\documentclass[a4paper,oneside,11pt]{article}
\usepackage{cprform}
\usepackage{amsmath,amstext,amsfonts,amsbsy,amssymb,amscd,epsfig,lscape}

\usepackage{epsfig,macros,cite}
\usepackage{amssymb}
\usepackage{subfigure}

\newcommand{\ci}[1]{\boldsymbol{#1}}

\newcommand{\Dslash}{\relax{\kern+.25em / \kern-.70em D}}

\newcommand{\Real}{\relax{\mathsf{\Gamma\kern-.35em R}}}
\newcommand{\Int}{\relax{\mathsf{Z\kern-.40em Z}}}

\newcommand{\obar}[1]{\kern3pt\overline{\kern-2pt #1\kern-0pt}\kern1pt}

\newcommand{\corrbar}[1]{\kern3pt\overline{\kern-2pt #1\kern-0pt}\kern1pt}

\newcommand{\oVApAVren}[1]{\kern3pt\overline{\kern-2pt #1\kern-0pt}\kern1pt_{\rm\scriptscriptstyle VA+AV;s}}

\newcommand{\zbar}{\kern3pt\overline{\kern-2pt Z\kern-0pt}\kern1pt}

\newcommand{\zbarVApAV}[1]{\kern3pt\overline{\kern-2pt Z\kern-0pt}\kern1pt_{\rm\scriptscriptstyle VA+AV #1}}

\begin{document}


\begin{titlepage}

\vspace*{-25truemm}
\begin{flushright}
\hspace*{-1cm}ROM2F/2010/15, RM3-TH/10-8, DESY 10-138, \\ ICCUB-10-052, UB-ECM-PF-10-028, IFIC/10-33, LPSC10137
 
\end{flushright}\vspace{5truemm}

\centerline{\Large \bf $\mathbf{B_K}$-parameter from $\mathbf{N_f = 2}$ twisted mass lattice QCD}
\vskip 5 true mm
\centerline{\bigrm M.~Constantinou$^a$, P.~Dimopoulos$^b$, R.~Frezzotti$^{cd}$, K. Jansen$^{e}$, V.~Gimenez$^f$,}
\centerline{\bigrm V.~Lubicz$^{gh}$, F.~Mescia$^i$, H.~Panagopoulos$^a$, M.~Papinutto$^{j}$, G.C.~Rossi$^{cd}$,}
\centerline{\bigrm S.~Simula$^h$, A.~Skouroupathis$^a$, F.~Stylianou$^a$, A.~Vladikas$^d$}

\vspace*{4truemm}
\begin{figure}[!h]
  \begin{center}
    \includegraphics[scale=0.90]{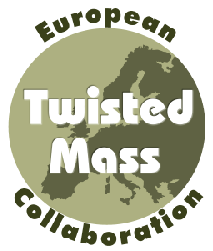}
 \end{center}
\end{figure}

\vskip 0 true mm
\centerline{\it $^a$ Department of Physics, University of Cyprus}
\centerline{\it P.O. Box 20537, Nicosia CY-1678, Cyprus}
\vskip 2 true mm
\centerline{\it $^b$ Dipartimento di Fisica, Sapienza, Universit\`a di Roma}
\centerline{\it Piazzale A. Moro, I-00185 Rome, Italy}
\vskip 2 true mm
\centerline{\it $^c$ Dipartimento di Fisica, Universit\`a di Roma ``Tor Vergata''}
\centerline{\it Via della Ricerca Scientifica 1, I-00133 Rome, Italy}
\vskip 2 true mm
\centerline{\it $^d$ INFN, Sezione di ``Tor Vergata"}
\centerline{\it c/o Dipartimento di Fisica, Universit\`a di Roma ``Tor Vergata''}
\centerline{\it Via della Ricerca Scientifica 1, I-00133 Rome, Italy}
\vskip 2 true mm
\centerline{\it $^e$ DESY, Zeuthen}
\centerline{\it Platanenallee 6, D-15738 Zeuthen, Germany}
\vskip 2 true mm
\centerline{\it $^f$  Departament de Fisica Te\`orica and IFIC, Univ. de Val\`encia-CSIC}
\centerline{\it Dr.~Moliner 50, E-46100 Val\`encia, Spain}
\vskip 2 true mm
\centerline{\it $^g$ Dipartimento di Fisica, Universit\`a di Roma Tre}
\centerline{\it Via della Vasca Navale 84, I-00146 Rome, Italy}
\vskip 2 true mm
\centerline{\it $^h$ INFN, Sezione di Roma Tre}
\centerline{\it c/o Dipartimento di Fisica, Universit\`a di Roma Tre}
\centerline{\it Via della Vasca Navale 84, I-00146 Rome, Italy}
\vskip 2 true mm
\centerline{\it $^i$ Departament d'Estructura i Constituents de la Mat\'eria}
\centerline{\it Universitat de Barcelona, 6a pianta, Diagonal 647}
\centerline{\it E-08028 Barcelona, Spain}
\vskip 2 true mm
\centerline{\it $^j$ Laboratoire de Physique Subatomique et de Cosmologie}
\centerline{\it UJF/CNRS-IN2P3/INPG}
\centerline{\it 53 rue des Martyrs, 38026 Grenoble, France}

\vskip 30 true mm


\thicktablerule
\vskip 3 true mm
\noindent{\tenbf Abstract}
\vskip 1 true mm
\noindent
{\tenrm  We present an unquenched $N_f=2$ lattice computation of the 
 $B_{K}$ parameter  
 which controls $K^0-\bar K^0$ oscillations. A partially quenched setup is employed with two maximally twisted 
dynamical (sea) light Wilson quarks, and valence quarks   
of both the maximally twisted  
and the Osterwalder--Seiler variety. Suitable combinations of these two kinds of valence quarks lead to a lattice 
definition of the $B_{K}$ parameter which is both multiplicatively renormalizable and O($a$) improved. 
Employing the non-perturbative RI-MOM scheme, 
in the continuum limit and at the physical value of the 
pion mass we get $B^{\rm RGI}_K=0.729\pm 0.030$, a number well in line with the existing quenched 
and unquenched determinations. 
}
\vskip 3 true mm
\thicktablerule
\eject
\end{titlepage}

\section{Introduction}
\label{sec:intro}

\begin{figure}[!ht]
\begin{center}
\includegraphics[scale=0.7,angle=-0]{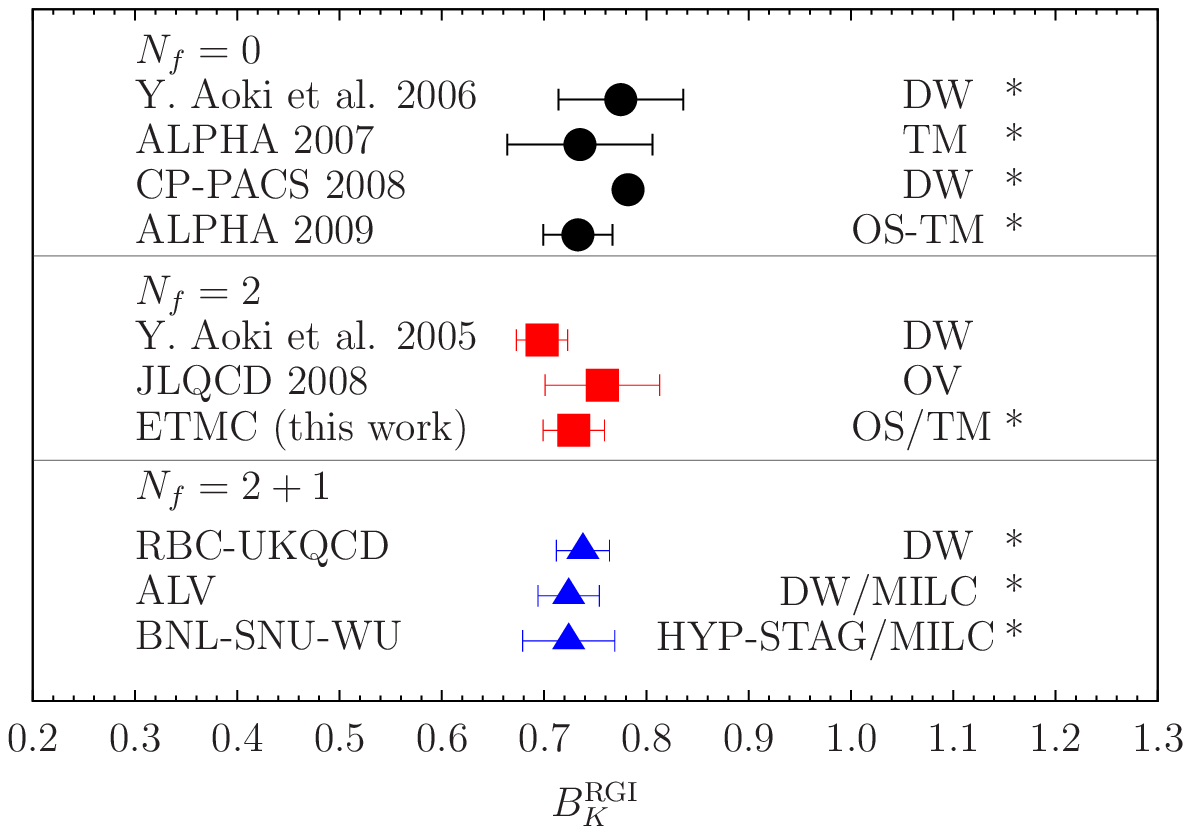}
\end{center}
\vspace{-1.cm}
\begin{center}
\caption{A compilation of quenched and unquenched results for $B_{K}^{\rm RGI}$. Data are from 
refs.~\cite{Aoki:2005ga,Dimopoulos:2007cn,Nakamura:2008xz,Dimopoulos:2009es,Aoki:2004ht,Aoki:2008ss,Kelly:2009fp,Aubin:2009jh,Bae:2010ki}, 
respectively. The symbol $^\star$ denotes determinations where the continuum extrapolation was carried out.}
\label{fig:figcomp}
\end{center}
\end{figure}

 Strong interaction effects in neutral $K^0-\bar K^0$ meson oscillations are parametrized by the 
so-called ``bag parameter''  $B_{K}$~\cite{Buras:1998raa}. Several lattice computations of $B_{K}$ based 
on different lattice fermion discretizations have been performed in the years, mostly in the quenched 
approximation. Attention has now shifted to unquenched estimates. See refs.~\cite{Lellouch:latt08,Lubicz:2010nx} 
for an updated review on the subject. A compilation of recent data is provided in Fig.~\ref{fig:figcomp} 
where renormalization group invariant (RGI) values of $B_{K}$ (denoted as $B_K^{\rm RGI}$ in this paper) are reported. 
Remarkably, no significant dependence of $B_K^{\rm RGI}$ on the number $N_f$ of sea quark flavours is observed.

The standard way of computing $B_{K}$ with Wilson fermions yields limited accuracy 
due to   
 partial control of two sources of systematic effects.

$\bullet$ First of all, the $\Delta S = 2$ four-fermion operator relevant for the $B_{\rm K}$ calculation, namely 
\begin{equation}
O^{\Delta S =2}=\bar s \gamma_\mu(1-\gamma_5) d\,\bar s \gamma_\mu(1-\gamma_5) d 
\label{DS2}
\end{equation}
mixes under renormalization with four other operators of the same dimension, but belonging to 
different chiral representations~\cite{Bochicchio:1985xa} (the so-called ``wrong chirality mixing'' 
phenomenon). The reason behind this feature is the lack of chiral symmetry of the 
regularized lattice action. In other regularizations (e.g.\ staggered or Ginsparg--Wilson fermions) 
partial or full chiral symmetry ensures that the operator~(\ref{DS2}) is multiplicatively renormalizable 
(i.e.\ does not mix with other operators). An important detail here is that mixing actually only 
affects the parity-even   
component of the four-fermion operator~(\ref{DS2}). In the following 
we will denote it 
by $O_{\rm VV+AA}$ dropping the superscript $\Delta S = 2$ for simplicity. 
Its parity-odd component,   
similarly called $O_{\rm VA+AV}$, is protected against mixing by discrete symmetries, irrespective 
of the status of chiral symmetry in the lattice action, as shown in~\cite{Bernard:1987pr,Donini:1999sf}.

$\bullet$  Secondly while a lattice $B_{K}$ estimate, based on staggered or Ginsparg--Wilson quarks, would be 
 affected by only O($a^2$) cutoff artifacts at finite gauge coupling, computations based on plain Wilson 
fermions suffer from O($a$) discretization errors. A well known remedy to this state of affairs is to make use of the 
Symanzik improvement strategy which entails the inclusion of the Clover term in the action, as well as a number of 
dimension-7 counter-terms in the operator. However the non-perturbative determination of the coefficients with which 
the counter-term operators should enter would greatly complicate the computation and increase the systematic error.

A  proposal to circumvent the first of these problems was put forward in ref.~\cite{Becirevic:2000cy} 
where by the use of Ward--Takahashi identities (WTIs) it was shown that the matrix element 
$\langle \bar K^0 \vert O_{\rm VV+AA}(0) \vert K^0 \rangle$ can be expressed in terms 
of the matrix element $\int d^4x \langle \bar K^0 \vert O_{\rm VA+AV}(0) P(x) \vert K^0 \rangle$,
where $P$ is the pseudoscalar quark density. It is seen that the latter involves the multiplicatively 
renormalizable parity-odd part of the four-fermion operator~(\ref{DS2}). The price to pay for this 
result is that now one has to compute a four-point correlation function (while normally $B_{K}$ 
is obtained from the large-time asymptotic limit of a three-point correlation function). This implies 
increased statistical fluctuations which in practice may offset the gain stemming from the absence 
of wrong chirality mixings~\cite{Becirevic:2004aj}.

Other attempts to avoid this problem have been tried out in refs.~\cite{Dimopoulos:2006dm,Dimopoulos:2007cn}, 
inspired  
by the twisted-mass formulation of lattice QCD (tm-LQCD)~\cite{Frezzotti:2000nk}. 
Two possibilities have been explored. 
 
The first one consists in using a lattice fermion action with a maximally twisted 
up-down fermion doublet and a standard (untwisted) Wilson strange quark. 
As a second possibility, viable  
only in the quenched approximation (for  lack of reality of the resulting fermion determinant), a degenerate 
down-strange doublet with twist angle $\pi/4$ is introduced. In both variants the change of variables that
brings the tm-LQCD quark action to the standard Wilson form happens to map the parity-even $\Delta S = 2$ operator
onto its parity-odd counterpart, thereby implying multiplicative renormalizability.

All the methods described above~\cite{Becirevic:2000cy,Dimopoulos:2006dm,Dimopoulos:2007cn}, while
avoiding the operator mixing problem, lead to $B_{K}$ estimates which suffer from O($a$) discretization 
effects. Achieving O($a$) improvement, without the use of Symanzik counter-terms, necessitates having all 
fermions at maximal twist (Mtm-LQCD~\cite{Frezzotti:2003ni}). Clearly, this must be done while preserving the 
mapping from parity-even to parity-odd $\Delta S = 2$ operator, so as not to loose multiplicative renormalizability
(see the discussion in ref.~\cite{tmqcd:DIrule}). 

This can actually be   
achieved by  using a different regularization 
for sea and valence quarks. The idea proposed 
in ref.~\cite{Frezzotti:2004wz} and adopted in the present study
is to have maximally twisted sea quarks in combination with Osterwalder--Seiler (OS) valence quarks. We recall  
that any OS fermion can be thought of as  a component of a maximally twisted doublet
with twist angle $+\pi/2$ or $-\pi/2$, depending on the value of its Wilson $r$-parameter.

Then the relevant $\Delta S=2$ four-fermion operator is  
defined in terms of four distinct (maximally twisted) 
valence flavours, three with twist angle, say, $+\pi/2$, and the fourth with twist angle $-\pi/2$. The discrete symmetries 
of the valence quark action combined with the structure of the resulting four-fermion operator ensure 
that the latter is multiplicatively renormalizable, while its $K^0 - \bar K^0$ matrix element 
remains automatically O($a$) improved~\cite{Frezzotti:2004wz}. The same is then true for $B_{K}$.

In the setup of ref.~\cite{Frezzotti:2004wz} unitarity violations occur as a consequence of the different  
regularization of sea and valence quarks. However, if renormalized masses of sea and valence fermions are matched, 
one can prove~\cite{Frezzotti:2004wz} that unitarity violations are mere O($a^2$) effects. Moreover, 
it just happens that the valence quark content of the $K^0$-meson consists of a strange/down pair with, say, 
the same twist angle, while the $\bar K^0$-meson has necessarily a quark-anti-quark pair with opposite twist angles. 
Thus the two pseudoscalar mesons have masses that at non-zero lattice spacing differ by (numerically important) O($a^2$)
effects, which mainly come from the lattice artifacts (non-vanishing in the chiral limit) of the $K^0$-meson mass. 
The $\bar K^0$-meson mass on the other hand exhibits only O($a^2(\mu_\ell+\mu_s)$) discretization errors 
(here $\mu_{\ell/s}$ denotes the mass of the light/strange quark). 
This effect has been recently studied in the quenched approximation 
adding the clover term in ref.~\cite{Dimopoulos:2009es},
where it is numerically demonstrated that the scaling behaviour of both $B_{K}$ and
 the decay constants of the 
$K^0$ and $\bar K^0$ mesons is only weakly affected (see also below). 
The situation about the O($a^2$) mass-splitting 
between $K^0$ and $\bar K^0$ is similar to the one already met in the unitary Mtm-LQCD framework, 
where a large O($a^2$) artefact is only observed in the mass of the neutral pseudoscalar meson as it is not 
associated to a conserved lattice axial current~\cite{Frezzotti:2007qv,Dimopoulos:2009qv}.

In this paper we present an unquenched computation of $B_{K}$ based on the strategy of  
ref.~\cite{Frezzotti:2004wz} which exploits the lattice data coming from the state-of-the-art 
tm-LQCD simulations carried out by the ETM Collaboration (ETMC) with $N_f=2$ dynamical 
(sea quark) flavours, at three lattice spacings and ``light" pseudoscalar meson masses in the
range 280~MeV $< m_{\rm PS} < $ 550~MeV. Renormalization is carried out non-perturbatively 
in the RI-MOM scheme~\cite{renorm_mom:paper1}. In the continuum limit and at the physical
value of the pion mass we get 
\begin{equation}
B_K^{\rm RGI} \; = \; 0.729(25)(17) \; 
\quad \Leftrightarrow \quad \; 0.729(30) \; \qquad @~N_f=2 \, ,
\label{BK-fin}
 \end{equation}
where in the first expression the two errors quoted are the statistical one (0.025)
and (our estimate of) the systematic uncertainty 
(0.017, resulting from the quadratic sum of
0.004 from renormalization, 0.009 from control/removal of cutoff effects and
0.014 from extrapolation/interpolation to the physical quark mass point).   
In the second expression the total error is obtained by a sum in quadrature of the statistical 
and systematic ones. The error budget is further discussed in sect.~\ref{sec:CBHAT}. 
The specification $@~N_f=2$ is to remind that the whole computation of $B_K^{\rm RGI}$ has
been carried out in QCD with two light ($u$ and $d$) dynamical quark flavours.
The quality of this result is extremely satisfactory and   
fully comparable with other quenched and unquenched numbers (see Fig.~\ref{fig:figcomp} 
for a compilation of recent determinations). 

The outline of the paper is as follows. In sect.~\ref{sec:tmQCD-gen} we briefly recall the theoretical 
framework which allows to obtain an O($a$) improved and multiplicatively renormalizable lattice 
expression of the $K^0-\bar K^0$ matrix element of the four-fermion $\Delta S =2$ operator~(\ref{DS2}). 
In sect.~\ref{sec:res} we give the details of our numerical analysis. Conclusions can be found in 
sect.~\ref{sec:concl}. We defer to a couple of Appendices a few more technical considerations. Preliminary 
reports about the present work have already appeared in refs.~\cite{BK-ETM-prel1,BK-ETM-prel2,Lubicz:2010nx}

\section{Twisted and Osterwalder--Seiler valence quarks at maximal twist}
\label{sec:tmQCD-gen}

The lattice setup of our study is that of ref.~\cite{Frezzotti:2004wz}. The regularization of sea 
quarks is different from that of valence quarks. The former is a standard tm-LQCD regularization with a doublet of 
degenerate, maximally twisted Wilson fermions (representing up and down flavours). In the so-called 
``physical'' basis, the Mtm-LQCD fermionic action takes the form 
\begin{equation}
S_{\rm Mtm} =  a^4 \sum_x \bar \psi (x) \Big \{ \frac{1}{2} \sum_\mu \gamma_\mu(\nabla_\mu 
+ \nabla^\ast_\mu ) - i \gamma_5 \tau^3 \big [ M_{\rm cr} - 
\dfrac{a}{2} 
\sum_\mu \nabla^\ast_\mu \nabla_\mu \big ] + \mu_{sea} \Big \} \psi(x)\, ,\label{SPHYS}
\end{equation}
where the Wilson's $r$ parameter has been set to unity,  
$\psi(x)$ is the quark flavour doublet, $\nabla_\mu$ and $\nabla_\mu^\ast$ are   
nearest-neighbour forward 
and backward lattice covariant derivatives, $\mu_{sea}$ is the (twisted) sea quark mass and $M_{\rm cr}$ the 
critical mass. The lattice pure gauge action is the tree-level improved action of ref.~\cite{Weisz:1982zw}, 
routinely adopted by the ETM Collaboration~\cite{Boucaud:2007uk} in studies with two light sea quark 
flavours. Valence quarks are regularized as Osterwalder--Seiler (OS) fermions~\cite{Osterwalder:1977pc}. 
The action of each OS valence flavour $q_f$ reads 
\begin{equation}
S_{\rm OS} =  a^4 \sum_x \bar q_f (x) \Big \{ \frac{1}{2} \sum_\mu \gamma_\mu(\nabla_\mu 
+ \nabla^\ast_\mu ) - i \gamma_5 r_f \big [ M_{\rm cr} - 
\frac{a}{2} \sum_\mu \nabla^\ast_\mu \nabla_\mu \big ] + \mu_f \Big \} q_f(x)\, ,
\label{OSvalact}
\end{equation}
where $aM_{\rm cr}$ is the same number as in eq.~(\ref{SPHYS}). The critical mass
$M_{\rm cr}$ is non-perturbatively tuned to its optimal value~\cite{Frezzotti:2005gi} 
as described in ref.~\cite{Boucaud:2007uk}. This ensures O($a$)-improvement of physical
observables and control of the leading chirally enhanced cutoff effects~\footnote{At the formal 
level a ghost action is also involved to cancel the fermionic determinant, arising from the integration 
over the $q_f$ and $\bar q_f$ valence degrees of freedom~\cite{Frezzotti:2004wz}.
The presence of ghosts has no practical consequences in the evaluation of correlation
functions of operators where no ghost fields appear.}.

\subsection{Lattice operators} \label{sec:LATOP}

For the computation of interest, it is convenient to introduce four OS valence quark flavours $q_1, q_2, q_3, q_4$. 
Eventually $q_1$ and $q_3$ will be identified with the strange quark by setting $\mu_1=\mu_3 \equiv \mu_h$, and $q_2$ and $q_4$ 
with the down quark by setting $\mu_2=\mu_4 \equiv \mu_\ell$. The key point is that the four-fermion operator
\begin{equation}
Q_{VV+AA} \,= \, 2 \left\{ [\bar q_1 \gamma_\mu q_2][\bar q_3 \gamma_\mu q_4] +
[\bar q_1 \gamma_\mu \gamma_5 q_2][\bar q_3 \gamma_\mu \gamma_5 q_4] + (q_2 \leftrightarrow q_4) \right\}
\label{Q_val}
\end{equation}
is multiplicatively renormalizable once the Wilson parameters appearing in the action of the OS fermions 
are taken to satisfy the relations 
\begin{equation}
r_1 = r_2 = r_3 = -r_4\, .
\label{RPAR}
\end{equation}
The factor 2 in the r.h.s.\ of eq.~(\ref{Q_val}) guarantees that the correlator~(\ref{PQP-correl}), where 
$Q_{VV+AA}$ is inserted between the interpolating $K^0=\bar q_2 q_1$ and $\bar K^0=\bar q_3 q_4$ fields,  
is normalized as in continuum QCD (as one can check by direct application of the Wick theorem). 
The proof of the multiplicative renormalizability of the operator $Q_{VV+AA}$ is based on the discrete 
symmetries of the OS action and is provided in full detail in ref.~\cite{Frezzotti:2004wz}. A simpler 
proof is readily obtained in the so-called ``twisted basis". In fact, upon performing the chiral field rotation
\begin{eqnarray}
&& q_f \to q_f^{tm}={\rm {e}}^{i\pi r_f \gamma_5/4}\,q_f \, ,\nonumber \\ 
&& \bar q_f \to \bar q_f^{tm} = \bar q_f {\rm {e}}^{i\pi r_f \gamma_5/4}
\label{CHIROT}
\end{eqnarray}
from the ``physical'' ($q_f$) to the ``twisted'' ($q_f^{tm}$) basis, 
the valence quark action in the massless limit (all $\mu_f$'s set to zero) takes the standard
(untwisted) Wilson form, while the parity-even operator $Q_{VV+AA}$ gets rotated onto the parity-odd 
$Q_{VA+AV}$. The latter, being expressed in terms of quark fields with standard Wilson action, 
is known~\cite{Bernard:1987pr,Donini:1999sf} to be multiplicatively renormalizable. 
Moreover, the argument implies that the renormalized operator is given by the formula
\begin{equation}
[Q_{VV+AA}]_{\rm R} \,= \, Z_{VA+AV} \, Q_{VV+AA}\label{QREN} \, ,
\end{equation}
where, for consistency with literature  
on Wilson fermions, the renormalization constant 
is named after the form ($VA+AV$) the operator takes in the ``twisted'' basis.	
In actual simulations we fixed the common value in eq.~(\ref{RPAR}) by setting $r_1=1$.

In order to compute $B_{K}$, we also need to consider the axial currents
\begin{equation}
A_\mu^{12} \,= \, \bar q_1 \gamma_\mu \gamma_5 q_2 \qquad \qquad
A_\mu^{34} \, = \, \bar q_3 \gamma_\mu \gamma_5 q_4 \label{ADEF}
\end{equation}
which, as a consequence of the relations between the Wilson parameters $r_1, \dots , r_4$ specified in
eq.~(\ref{RPAR}), when passing from the ``physical'' to the ``twisted'' basis take the form of axial and 
vector operators, respectively. Consequently they renormalize according to the relations 
\begin{equation}
[A_\mu^{12}]_{\rm R} \,= Z_A \, A_\mu^{12}  \qquad \qquad
[A_\mu^{34}]_{\rm R} \, = Z_V \, A_\mu^{34} \, .\label{AREN}
\end{equation}

\subsection{Lattice correlation functions} \label{sec:LATCO}

Next we proceed to construct the correlation functions, involving the operators that are necessary  
to  evaluate $B_{K}$.  The lattice is taken to be of size $L^3 \cdot T$, with  
periodic boundary conditions 
on all fields and   
in all directions, except for quark fields in the time direction which  
satisfy antiperiodic
boundary conditions. At a reference time slice $y_0$, we define a ``$K$-meson wall" 
with pseudoscalar quantum numbers and $\bar q_2$ and $q_1$ quark fields, namely
\begin{equation}
W^{21}(y_0) =   \Big{(}\dfrac{a}{L}\Big{)}^3\, \sum_{\vec y} 
\bar q_2(\vec y, y_0) \gamma_5 q_1(\vec y , y_0) \, .
\label{K-WALL}
\end{equation}
A second ``$K$-meson wall", $W^{43}(y_0+T/2)$, with $\bar q_4$ and $q_3$ quark fields,
is placed at the time slice $y_0 + T/2$. The four-fermion operator, 
$Q_{VV+AA}$, is inserted at position $x=(\vec x, x_0)$, with $x_0$ in the range $y_0 \ll x_0 \ll y_0+T/2$. In this way 
lattice estimators of the bare $B_{K}$-parameter can be calculated at several values of $x_0$ from the ratio
\begin{equation}
R(x_0) \,= \, \dfrac{C_3(x_0)}{C_2(x_0) \,\, C_2^\prime(x_0)}
\label{BKratio}
\end{equation}
involving the correlation functions~\footnote{In order to simplify   
the notation, we do not display the dependence 
on the reference time slice $y_0$. The latter is chosen randomly on each gauge configuration so as  
to reduce the 
magnitude of the autocorrelation time of $B_{K}$ lattice estimators.}  
\begin{eqnarray}
&&C_3(x_0) = \Big{(}\dfrac{a}{L}\Big{)}^3\sum_{\vec x} \langle W^{43}(y_0 + 
\frac{T}{2}) \,Q_{VV+AA} (x) \, W^{21}(y_0) \rangle\, ,
\label{PQP-correl}\\
&&C_2(x_0) = \Big{(}\dfrac{a}{L}\Big{)}^3\sum_{\vec x} \langle A_0^{12}(x) \,W^{21}(y_0) \rangle\, ,
\label{PP-12}\\
&&C_2^\prime(x_0) = \Big{(}\dfrac{a}{L}\Big{)}^3 \sum_{\vec x} \langle W^{43}(y_0 + \frac{T}{2})  \,A_0^{34}(x) \rangle \, .
\label{PP-34}
\end{eqnarray}

The signal-to-noise ratio is greatly enhanced by summing over the spatial position of the quark
and antiquark fields in each ``$K$-meson wall" as well as summing over that of the four-fermion operator.   
The $x_0$-behaviour of the $B_{K}$-estimator  excludes significant contaminations from 
excited states (see also sect.~\ref{sec:ERLD}). On the other hand in order   
to reduce the  cost  of computing the quark propagators we chose to
employ  
a stochastic method.  This consists in computing quark propagators 
by inverting the lattice Dirac operator 
for the relevant OS valence quark flavours on random sources of the form
\begin{equation}
\delta_{\alpha\gamma}  \delta_{x_0,z_0}
\delta_{ik}  \delta_{\vec{x},\vec{z}} \,\eta^{(z_0)}(k, \vec z) \, ,
\quad \gamma=1,2,3,4  \, , \quad z_0 = y_0, \, y_0 + T/2 \, ,
\label{Z2SOURCE}
\end{equation}
with free indices $\alpha$ (spin), $x_0$ (time), $i$ (colour),
$\vec{x}$ (space), while $\gamma$ and $z_0$ are fixed as indicated.
The $Z_2$-valued $\eta$ vectors carry only colour
and three-space indices (colour and three-space ``dilution'')
and are normalized according to
\begin{equation}
\langle \eta^{(z_0)}(k, \vec z\,) \eta^{(z_0)}(k', \vec z\,') \rangle
= \delta_{kk'} \delta_{\vec z \,\vec z\,'} \, .\label{NORMSOUR}
\end{equation}
Given the pattern of Wilson $r$-parameters specified in eq.~(\ref{RPAR}) and the 
invariance under $\gamma_5$-hermitian conjugation combined with $r_f \to -r_f$ of the
OS lattice Dirac operator, it is enough to compute for each gauge configuration the
propagator of the fields $q_1$ and $\bar q_2$ on the random source at $y_0$, and 
the propagator of the fields $q_3$ and $q_4$ on the random source at $y_0+T/2$. 
The conditions~(\ref{NORMSOUR}) 
guarantee that unbiased estimators of the correlators $C_3$, $C_2$ and $C_2^\prime$ are 
obtained~\footnote{This setup  
 was first presented in~\cite{BK-ETM-prel1}.}.   
We found that, for an ensemble of a few hundred gauge configurations, employing one
set of $Z_2$-valued random sources (namely those in eq.~(\ref{Z2SOURCE})) per gauge configuration 
is sufficient to achieve a good statistical precision for our $B_K$-estimators. 

\subsection{Lattice estimator of (renormalized) $B_K$} \label{sec:LATBK}

As we said, the masses  of (up/down) sea quarks and the masses $[\mu_2]_{\rm R}$, 
$[\mu_4]_{\rm R}$ of (down) valence quarks must be tuned to the same (physical) value. 
At the same time the valence mass parameters
$[\mu_1]_{\rm R}$ and $[\mu_3]_{\rm R}$ should be adjusted to the strange quark 
mass in order for the external states to be identified with $K$-particles~\footnote{We recall that,
since the quark mass renormalization constant of OS valence quarks is independent of the sign of the
corresponding Wilson $r$-parameter~\cite{Frezzotti:2004wz}, at the level of bare masses we must
simply require  $\mu = \mu_2 = \mu_4$ and $\mu_1 = \mu_3$.}. 
In this way the renormalized ratio $\hat{R}(x_0) \equiv \dfrac{Z_{VA+AV}}{Z_A Z_V} \,\, R(x_0) $
computed with distinct OS valence flavours in the partially quenched set up specified above, 
will differ from its continuum limit by only O($a^2$) discretization errors. 

One peculiar feature of our approach which we have to keep in mind in the analysis of simulation data 
is the fact that kaon and anti-kaon masses are not degenerate as the two mesons involve 
differently regularized valence quarks. In our setting the kaon is made up of the valence quarks 
$q_2\to q_d$ and $q_1\to q_s$ with $r_1=r_2$, while the anti-kaon consists of the valence quarks 
$q_4\to q_d$ and $q_3\to q_s$ with $r_3=-r_4$. The $K^0-\bar K^0$ mass difference 
$M^{12}-M^{34}\to M_K-M_{\bar K}$ is an O($a^2$) effect which has an origin similar 
to that of the well known mass splitting one encounters in the pion sector~\cite{Dimopoulos:2009qv}. 
Naturally the relative effect is much less important in the kaon case as the kaon mass is substantially larger 
than the pion mass~\footnote{The scaling of the $K^0-\bar K^0$ mass splitting has been studied 
in detail in the quenched approximation in ref.~\cite{Dimopoulos:2009es}.}.
 
Taking this mass difference into account, we write for the three-point correlation function the 
large time expansion ($y_0 \ll x_0 \ll y_0+T/2$) 
\begin{eqnarray}
\hspace{-1.2cm} L^6 C_3(x_0) & 
  \rightarrow & \langle 0 \vert W^{43} (0) \vert P^{34} \rangle \,
\langle P^{34} \vert Q_{VV+AA} (0) \vert P^{21} \rangle \, \langle P^{21} \vert W^{21} (0) \vert 0  \rangle 
\nonumber  \\
\hspace{-1.2cm}  & \times & \dfrac{1}{4M^{12}M^{34}} \exp[- M^{12}(x_0 - y_0)] \, \exp[- M^{34}(y_0+\frac{T}{2}-x_0)] \, .\label{C3EXP}
\end{eqnarray}
In this expression $\vert P^{ij} \rangle$ denotes a zero three-momentum pseudoscalar state with quark content 
$q_i$ and $\bar q_j$ ($i,j = 1,\cdots , 4$) and mass $M^{ij}$. Similarly we have
\begin{eqnarray}
\hspace{-1.5cm}&& L^3C_2(x_0) \, \rightarrow\,   \langle 0 \vert A_0^{12}(0) \vert P^{21} \rangle \,
\langle P^{21} \vert W^{21} (0) \vert 0 \rangle \, \dfrac{1}{2M^{12}} \exp[- M^{12}(x_0 - y_0)] \, ,\label{C12EXP}
\\
\hspace{-1.5cm}&& L^3 C_2^\prime(x_0) \, \rightarrow \,  \langle 0 \vert W^{43} (0) \vert P^{34} \rangle \,
\langle  P^{34} \vert  A_0^{34} (0) \vert 0 \rangle \,\dfrac{1}{2M^{34}}  \exp[- M^{34}(y_0+\frac{T}{2}-x_0)] \, . \label{C34EXP}
\end{eqnarray}
Furthermore, defining the pseudoscalar decay constants $F^{ij}$ through the equations 
\begin{eqnarray}
 \langle 0 \vert A_0^{12}(0) \vert P^{21} \rangle \, = \, M^{12}F^{12}  
\qquad \qquad
\langle P^{34} \vert A_0^{34}(0) \vert 0 \rangle \,= \, M^{34}F^{34} \, ,
\end{eqnarray}
one finds that they are not equal, but again they differ by O($a^2$) effects. 

The renormalized lattice $B_{K}$ parameter which we will have to extrapolate to the 
continuum limit is finally obtained by averaging over the $x_0$-plateau 
(which lies between $y_0$ and $y_0+T/2$) of the estimator $\hat{R}(x_0)$, viz. 
\begin{equation}
\dfrac{8}{3} \hat{B}_{K,{\rm lat}}
\Leftarrow  \hat{R}(x_0) \equiv
\dfrac{Z_{VA+AV}}{Z_A Z_V} \,\, R(x_0)=  
\dfrac{Z_{VA+AV}}{Z_A Z_V} 
\dfrac{\langle P^{34} \vert Q_{VV+AA} (0) \vert P^{21} \rangle}
{M^{12}F^{12}M^{34}F^{34}}\, ,\label{BKR}
\end{equation} 
where the exponentials in eqs.~(\ref{C3EXP}), (\ref{C12EXP}) and~(\ref{C34EXP}) exactly cancel out 
and the numerical factor $8/3$ in the l.h.s.\ of this equation is there for conventional reasons, i.e.\  
it ensures that $B_K$ represents the four-fermion matrix element in units of
the vacuum saturation approximation
value. 

\subsection{Scaling checks} \label{sec:SCACH}

In this section we present some tests aimed at checking the size of the discretization errors affecting 
the quantities that enter our computation of $B_K$ (i.e.\ those that appear in eq.~(\ref{BKR})). In order to
keep the issue of lattice artifacts separated from those related to the extrapolation to ``physical quark mass point'', 
the present scaling tests have been performed at fixed reference values of the  
renormalized ``light'' and ``heavy'' (would be strange) quark masses in the 
$\overline{\rm MS}$-scheme at the scale of 2~GeV. 
For convenience we take  $\hat\mu_\ell^* \sim 40$~MeV and $\hat\mu_h^* \sim 90$~MeV. 

\begin{figure}[!ht]
\begin{center}
\subfigure[]{\includegraphics[scale=0.50,angle=-0]{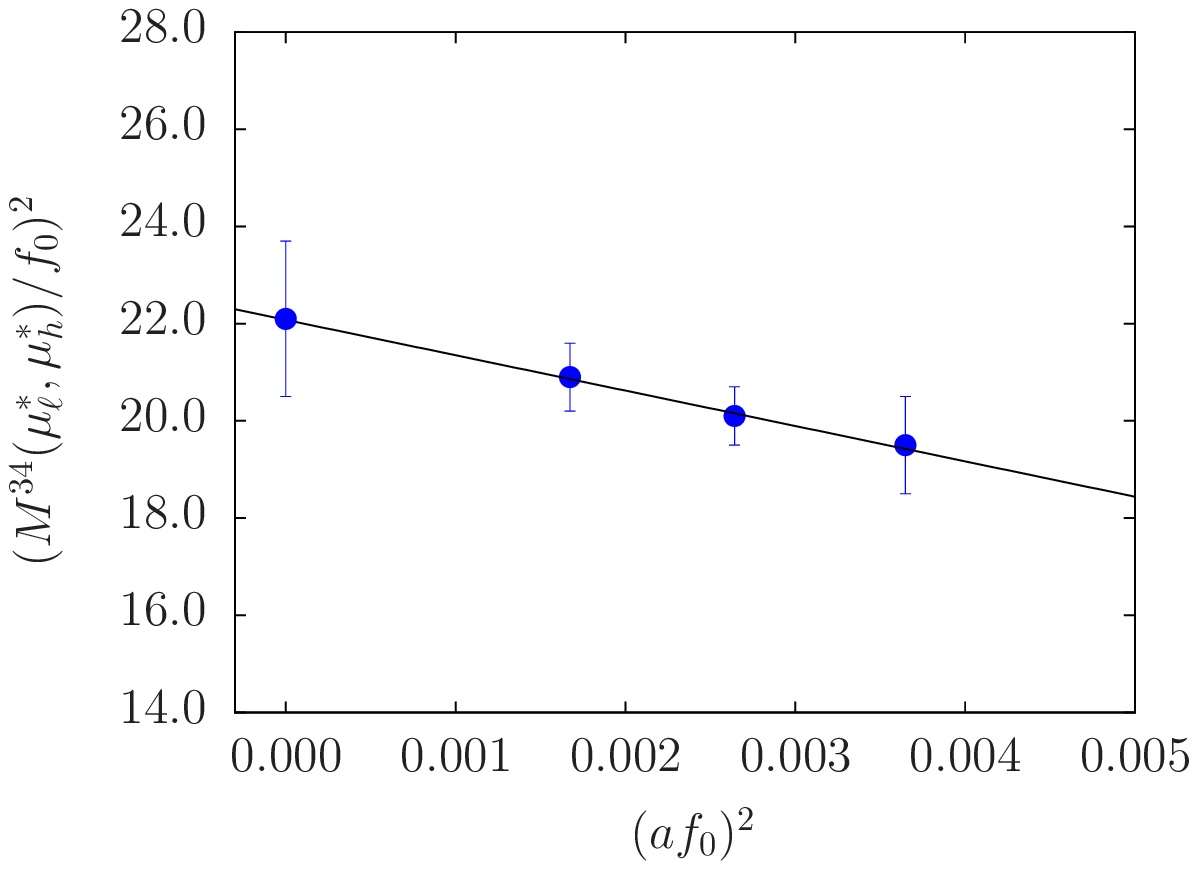}} 
\subfigure[]{\includegraphics[scale=0.50,angle=-0]{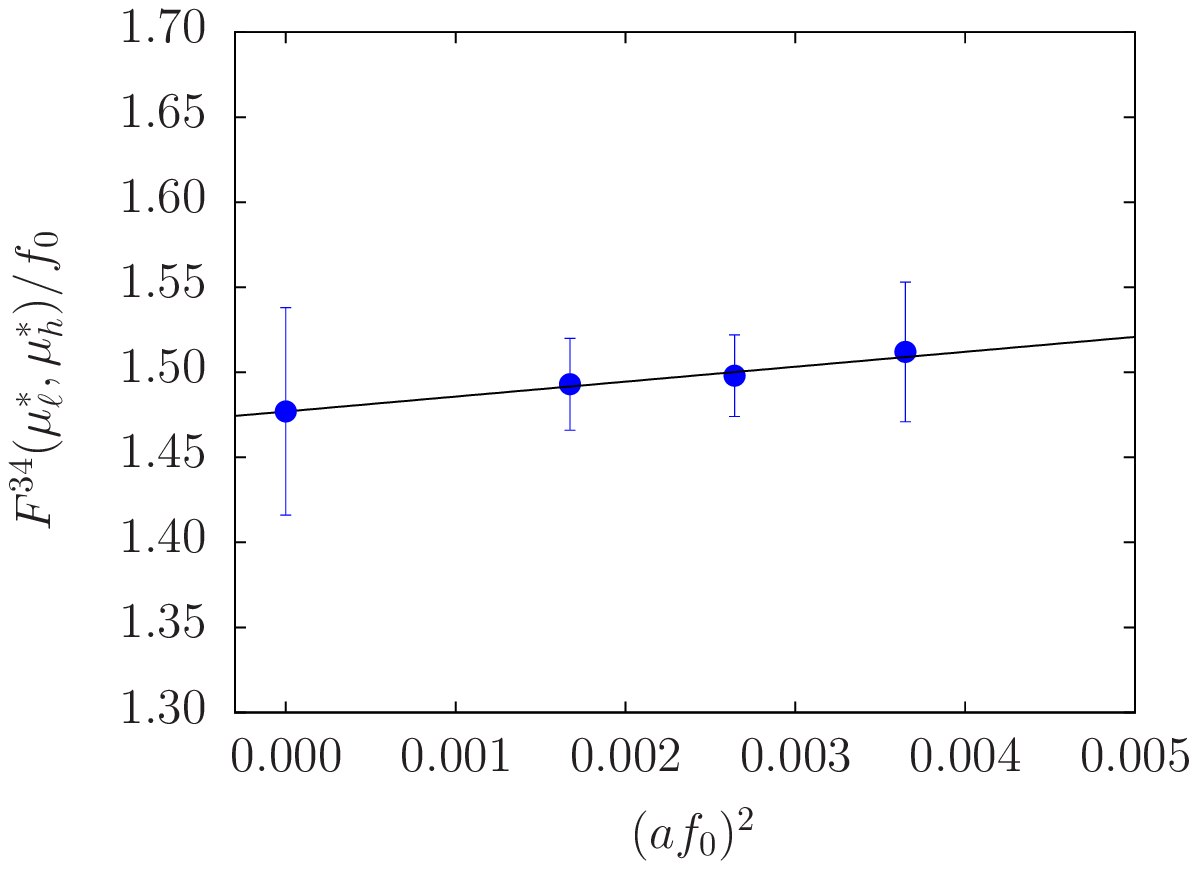}} 
\subfigure[]{\includegraphics[scale=0.50,angle=-0]{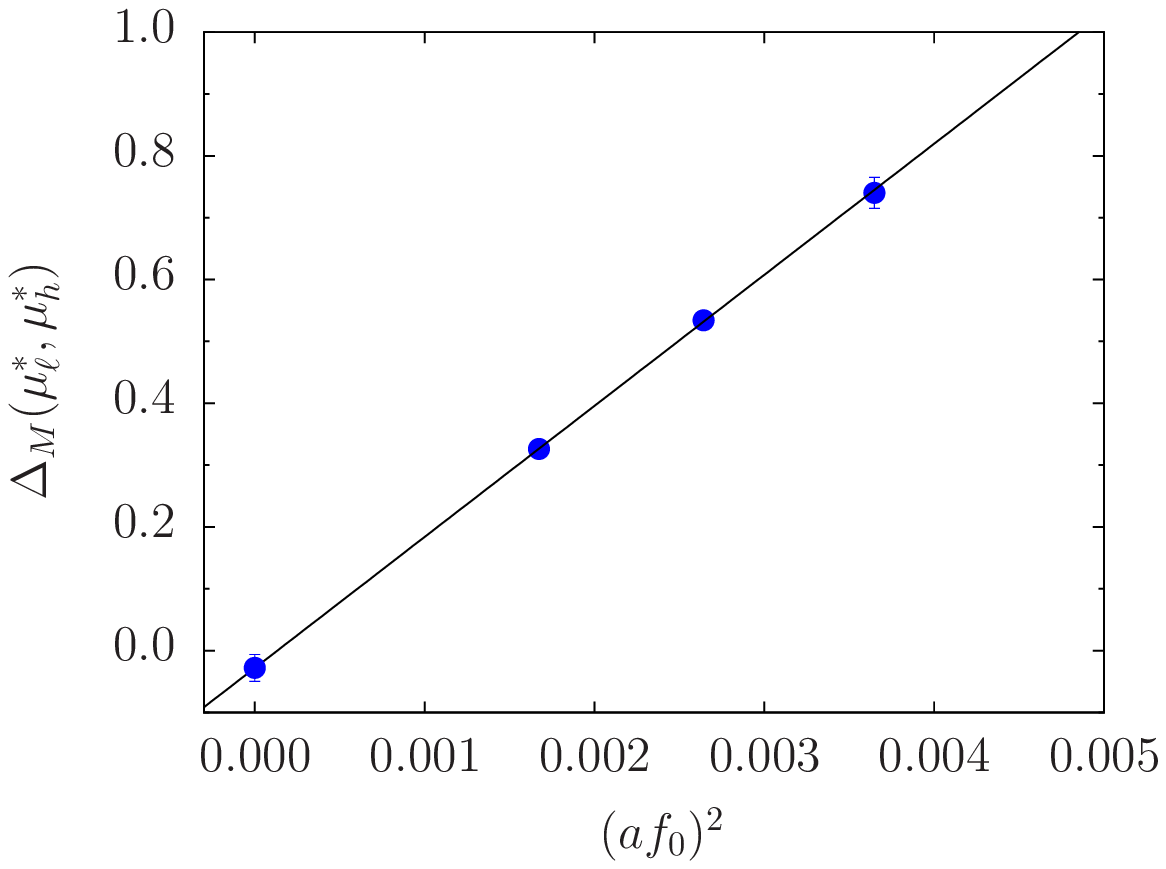}} 
\subfigure[]{\includegraphics[scale=0.50,angle=-0]{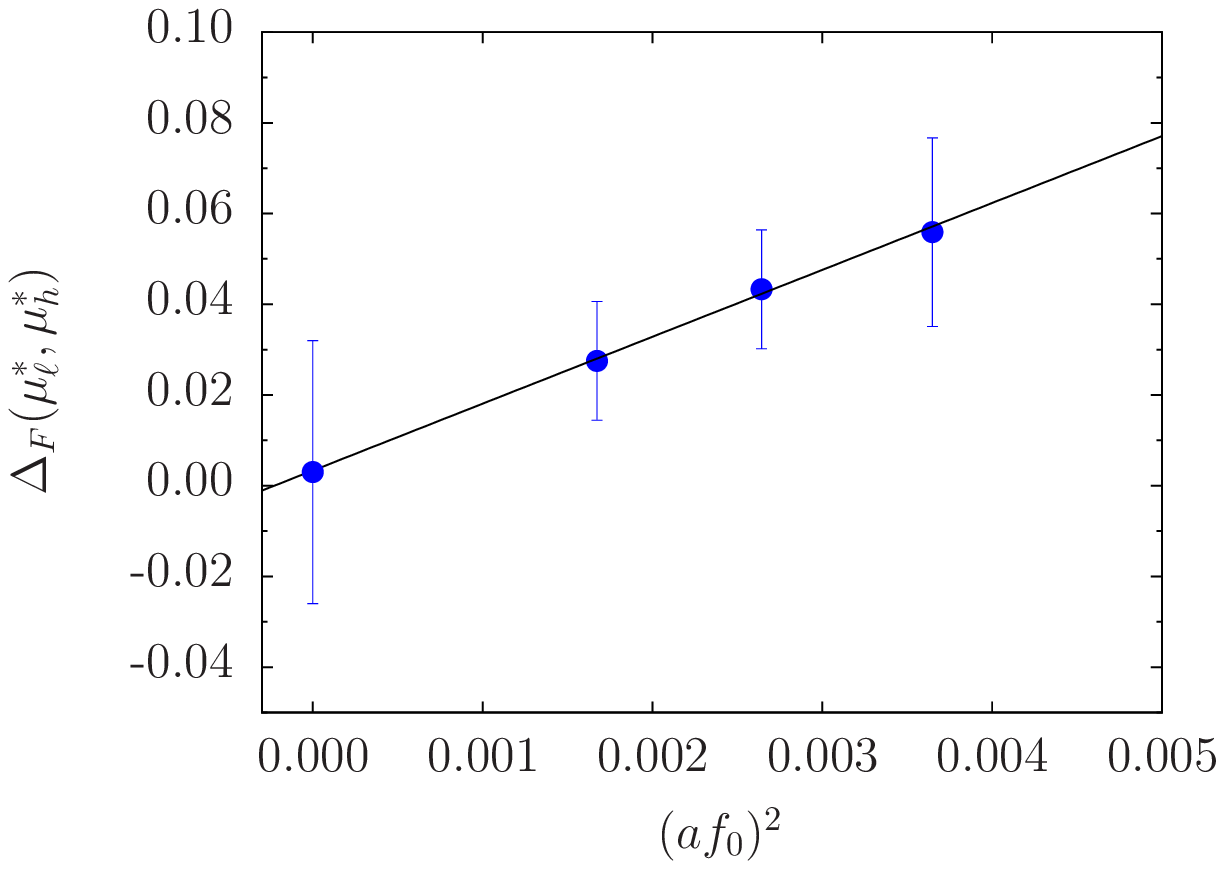}} 
\end{center}
\vskip -1.5cm
\begin{center}
\vspace{-0.5cm}
\caption{Data for
$(M^{34}/f_0)^2$ (panel (a)), $F^{34}/f_0$ (panel (b)), 
$\Delta_M$ (panel (c)) and $\Delta_F$ (panel (d)) as functions of $(af_0)^2$ at
the reference quark masses $\hat\mu_\ell^* \sim 40$~MeV and $\hat\mu_h^* \sim 90$~MeV
(in the $\overline{\rm MS}$ scheme at 2~GeV).
We display the best linear fit in $(af_0)^2$ and the corresponding continuum limit result.
}
\label{fig:DMAQ}
\end{center}
\end{figure}

In Fig.~\ref{fig:DMAQ}, panels (a) and (b), we show the scaling behaviour of $(M^{34}/f_0)^2$ and
$F^{34}/f_0$, i.e.\ the mass squared and the decay constant of the ``kaon'' made up of valence 
quarks $\bar q_4$ and $q_3$, normalized by the chiral limit pion decay constant $f_0 \simeq
121$~MeV determined in ref.~\cite{Baron:2009wt}. 
Owing to the choice $r_3 =-r_4$ the quantities $M^{34}$ and $f^{34}$ are our lattice estimators
of the ``kaon'' mass and decay constant that should exhibit best scaling properties. The plots
in panels (a) and (b) fully confirm this expectation, showing a nice $a^2$-scaling with
cutoff effects for the lattice ``kaon'' mass and decay constant estimates that are at
the level of few percents (from the largest lattice spacing to the continuum limit).

In Fig.~\ref{fig:DMAQ}, panels (c) and (d), we show instead the scaling behaviour of the 
difference of the ``kaon'' mass squared and decay constant lattice estimators entering
our $B_K$ computation, namely
\begin{equation}
\Delta_M=\dfrac{ (M^{12})^2-(M^{34})^2 }{(M^{34})^2} \, , \qquad
\Delta_F= - \dfrac{F^{12}-F^{34}}{F^{34}}\, . \label{DeltaMF} 
\end{equation}
The quantities~(\ref{DeltaMF}) are by construction mere O($a^2$) lattice artifacts.
We see from the panels (c) and (d) that both $\Delta_M$ and $\Delta_F$ 
extrapolate nicely to zero in the continuum limit, as expected.  
The large values of $\Delta_M$  in Fig.~\ref{fig:DMAQ}(c)
signal the presence of a substantial cutoff effect on $M^{12}$, with
$M^{12}$ on the coarsest lattice differing by about 30\% from its continuum
limit value ($\sim 570$~MeV).  
On the other hand, the lattice artifact difference between $F^{12}$ at
the largest lattice spacing and its continuum limit value is only about 5\%. This
is readily inferred by comparing Fig.~\ref{fig:DMAQ}(b) and Fig.~\ref{fig:DMAQ}(d).
  
Having taken $r_1=r_2$, the large and positive lattice artifact observed in $M^{12}$ 
was expected   
since for  OS valence quark doublets, 
(unlike the case of twisted-mass quark pairs) 
no isospin component of the lattice isotriplet 
axial current is conserved in the limit of vanishing quark mass.
Relying on arguments similar to those given in ref.~\cite{Dimopoulos:2009qv}, 
one can however conjecture that the large discretization error in $M^{12}$ 
is due to dynamically large matrix elements entering the Symanzik description
of the lattice OS pseudoscalar meson mass, rather than to large coefficients in front
of some terms of the Symanzik local effective action for 
OS valence fermions~\footnote{When extending the analysis of ref.~\cite{Dimopoulos:2009qv}
to the OS doublet case one finds that, in contrast   
to the situation in the twisted mass doublet case, 
matrix elements with both one insertion of the $d=6$ Symanzik local effective action term and 
two insertions of the $d=5$ term contribute to the cutoff effects in $(M^{12})^2$.}.
This conjecture suggests that the large cutoff effect detected in the lattice OS pseudoscalar meson mass is
peculiar to this quantity (or to others directly related to it) but not a general feature of physical observables built
in terms of OS valence quarks. The small cutoff effects observed in $F^{12}$ is well in line with such an expectation. 

\vskip 0.2cm
\begin{figure}[!ht]
\begin{center}
\includegraphics[scale=0.7,angle=-0]{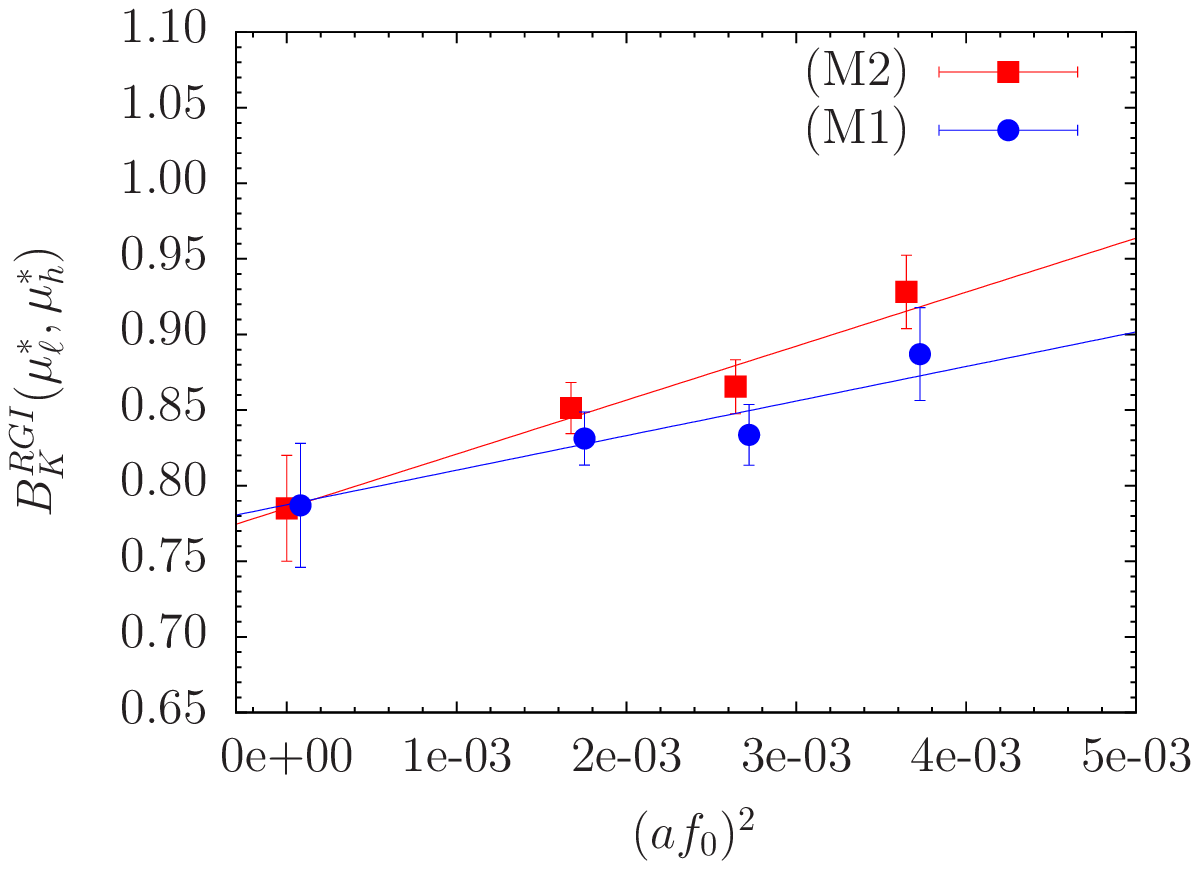}
\end{center}
\begin{center}
\vspace{-0.5cm}
\caption{$B_{K,{\rm lat}}^{\rm RGI}$ as function of $(af_0)^2$, at the same
reference quark masses, $\hat\mu_\ell^*$ and $\hat\mu_h^*$, as in Fig.~\ref{fig:DMAQ}, 
is shown together with the best fit linear in $a^2$ and the corresponding continuum limit result. 
The two sets of data, labelled as M1 and M2, come from different procedures for evaluating
the renormalization constants $Z_{AV+VA}$ and $Z_A$, expected to agree in the continuum limit.}
\label{fig:BKscal}
\end{center}
\end{figure}

In Fig.~\ref{fig:BKscal} we show the scaling behaviour of
$B_{K,{\rm lat}}^{\rm RGI}(\hat\mu_\ell^*, \hat\mu_h^*)$ evaluated according to eq.~(\ref{BKR})
using two different methods (referred to as M1 and M2) 
for the evaluation of the renormalization constants $Z_{VA+AV}^{\rm RGI}$ and $Z_A$. 
The two methods (discussed in sect.~\ref{sec:ZAVVA} and Appendix~A) differ only 
in the way one deals with O($a^2$) artifacts and other unwanted systematic effects pertaining
to the RI-MOM scheme computation of renormalization constants. The somewhat different slope in $a^2$ 
of the data-points in Fig.~\ref{fig:BKscal} should be ascribed to these effects. Nevertheless in both cases 
the $a^2$-scaling behaviour of $B_{K,{\rm lat}}^{\rm RGI}(\hat\mu_\ell^*, \hat\mu_h^*)$ 
is very good and the   
extrapolated continuum limit values agree very well. Cutoff artifacts in 
$B_{K,{\rm lat}}^{\rm RGI}(\hat\mu_\ell^*, \hat\mu_h^*)$ are about 10\% with the method M1 
and a bit larger if the method M2 is adopted. 

Two remarks are in order here. First of all, our choice of the lattice normalization
factor $(M^{12}F^{12} M^{34}F^{34})^{-1}$ in the r.h.s.\ of eq.~(\ref{BKR}) turns out
to be very beneficial   
as it partially compensates  the discretization errors coming 
from the matrix element 
$\langle P^{34} \vert Q_{VV+AA} (0) \vert P^{21} \rangle $.
Secondly, the cutoff effects in the RI-MOM scheme computation of $Z_A$ and $Z_{VA+AV}$
are significantly reduced  
when  1-loop perturbative lattice artifacts (i.e.\ O($a^2 g_0^2$) 
effects) are subtracted. The latter have been computed in refs.~\cite{Constantinou:2010gr,HP4ferm}.

In conclusion, in spite of the occurrence of a substantial O($a^2$) artifact in the mass
of the lattice state $\vert P^{21}\rangle$ and the ensuing fact that the 
$\langle P^{34} \vert Q_{VV+AA} (0) \vert P^{21} \rangle$ matrix element is evaluated with an
O($a^2$) four-momentum transfer, the scaling of 
$B_{K,{\rm lat}}^{\rm RGI}(\hat\mu_\ell^*, \hat\mu_h^*)$
with $a^2$ is found to be well under control, thereby allowing for a reliable continuum limit
extrapolation. The same is thus expected, and found to be true (see sect.~\ref{sec:CCE}), also upon
approaching the physical quark mass point.

\section{Simulations, data analysis and results}
\label{sec:res}

The ETM Collaboration has generated $N_f =  2$ configuration 
ensembles at three values of the inverse bare gauge 
coupling ($\beta$) and at a number of light quark masses ($\mu_{sea}$). 
Several quantities entering the data analysis of this paper, such as
the charged pion mass $aM_{\ell\ell}^{\rm tm}$~\footnote{We use this notation to make clear
that this is the mass of the ``charged'' pion which is known to be
affected by only small O($a^2\mu_\ell$, $a^4$) lattice
artifacts~\cite{Frezzotti:2005gi,Frezzotti:2007qv,Dimopoulos:2009qv}.},
the low-energy constants $af_0$ and $a\hat B_0$ and the chiral limit value of $a/r_0$,
are derived from ETMC data~\cite{Boucaud:2007uk,Boucaud:2008xu,Baron:2009wt}.
In this paper the  
symbol  $\hat Q$ means that the quantity  
$Q$ is renormalized in the $\overline{\rm MS}$ scheme at the 2~GeV scale.
We note that, since in our analysis the continuum limit value 
$\hat B_0 = Z_P B_0$ is employed, in order to evaluate 
the RGI quantity $B_0\mu_\ell = \hat B_0 \hat\mu_\ell $
one must know $\hat\mu_\ell = Z_P^{-1}\mu_\ell$ at the $\beta$-values of interest.
The necessary renormalization constants of quark bilinears,
such as $Z_P$, $Z_A$ and $Z_V$, have been computed non-perturbatively 
in the RI-MOM scheme~\cite{renorm_mom:paper1} 
as reported in ref.~\cite{Constantinou:2010gr} and converted to 
the $\overline{\rm MS}$ wherever necessary. 
The calculation of the four-fermion operator renormalization constant is new 
and discussed below in subsect.~\ref{sec:ZAVVA}.

\subsection{Extracting bare $B_K$ estimates from lattice data}
\label{sec:ERLD}

In the present computation we deal with $h\ell$ pseudoscalar mesons (``kaons'')
made of two valence quarks, $q_h$ and $q_\ell$, with bare masses $(\mu_h,\mu_\ell)$. 
The values of the mass parameters $(a\mu_h, a\mu_\ell)$ are chosen so as to have 
$\ell\ell$ pseudoscalar mesons (charged ``pions'' with $r_{u}=-\r_d$) with mass in 
the range 280-550~MeV and $h\ell$ pseudoscalar mesons with mass in the range 450-700~MeV, 
depending also (see sect.~\ref{sec:tmQCD-gen}) on whether the lattice mass
$M_{34} \leftrightarrow M_{\bar K}$, or rather $M_{12} \leftrightarrow M_K$,
is considered. The value of the light quark mass parameter, $a\mu_\ell$,
is common for sea and valence light (the would-be $u/d$) quark flavours, while 
the heavier quark (the would-be $s$) is quenched.
With this choice of mass parameters we will eventually be able,
as discussed in sect.~\ref{sec:CBHAT}, to make contact
with the physical $K^0$ and $\bar K^0$ mesons by interpolating our results
in $\mu_h \to \mu_s$ and extrapolating them in the $\mu_\ell \to \mu_{u/d}$
and continuum limit.

Simulation details are gathered in Tables~\ref{Table:bare380}, \ref{Table:bare390}, 
and~\ref{Table:bare405}, where we report, at each gauge coupling and for all the
available $(\mu_h,\mu_\ell)$-combinations, the values (in lattice
units) of the measured pseudoscalar meson masses and decay constants as well as 
the bare $B_K$-parameter. 
The latter (up to a trivial factor 8/3, see eq.~(\ref{BKR})) is denoted by $R$ 
and it is obtained by averaging over a plateau in $\tau \equiv x_0 - y_0$
of the lattice estimator $R(x_0)$ given in eq.~(\ref{BKratio}). 
In the table captions we specify the plateaux intervals  
$[(x_0-y_0)^{\rm min}/a,(x_0-y_0)^{\rm max}/a]$.
Indeed, the quantity $R(x_0)$ must be evaluated at large time separations.
In our case this means requiring that $x_0$ (the time coordinate where the four-fermion 
operator is located) is sufficiently far away from $y_0+ T/2 $ and $y_0$ 
(with $y_0$, $y_0+ T/2 $ the time coordinates of the two ``K-meson walls'', 
see sect.~\ref{sec:LATCO}). In this way 
the $K^0$- and the $\bar K^0$-state dominate the three-point correlator $C_3(x_0)$. 
Actually, due to the finite $T$ extension of the lattice, it is also necessary
that the contribution to the quantum-mechanical representation of $C_3(x_0)$  
from states wrapping around-the-world, such as that from $|\pi\pi \rangle$
or the $a^2$-suppressed one from $|\pi^0\rangle$, be negligible  
compared to the contribution from the vacuum state. For this purpose it is
important that $T$ is large enough with respect to the inverse lattice pion mass(es).

\begin{figure}[!ht]
\subfigure[]{\includegraphics[scale=0.50,angle=-0]{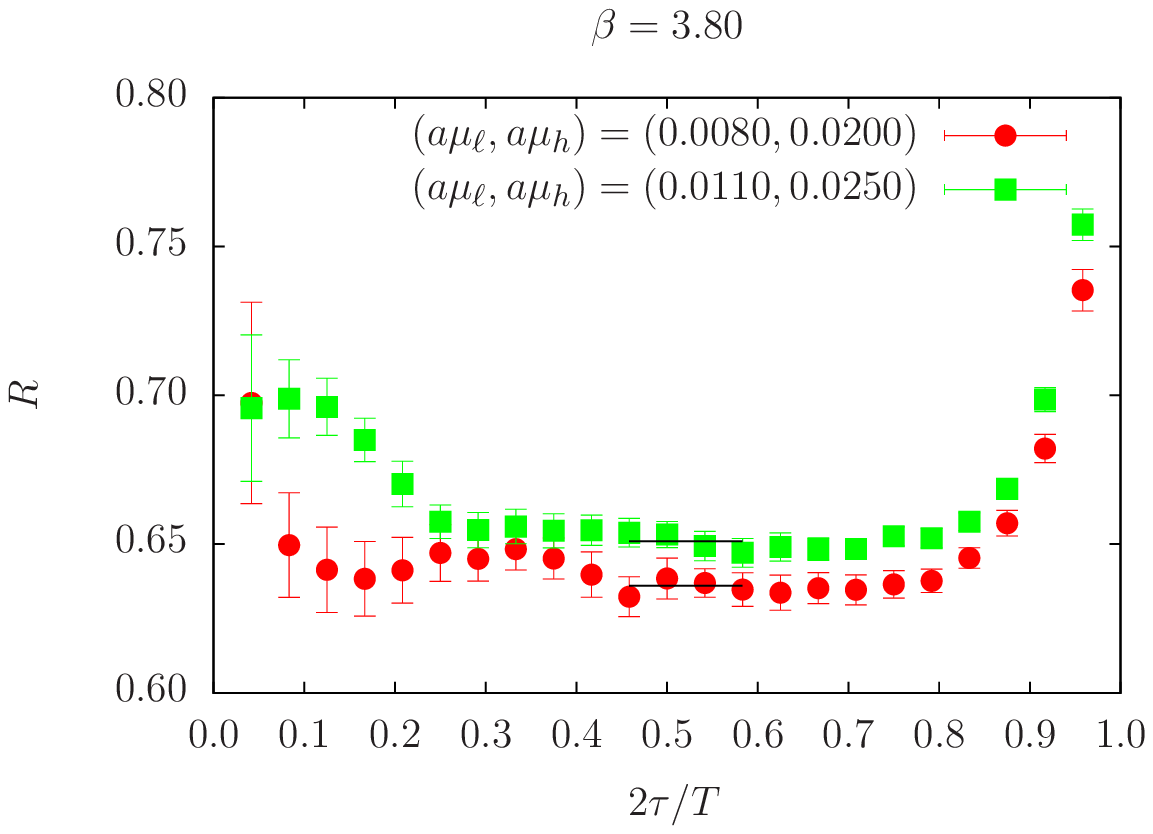}}
\subfigure[]{\includegraphics[scale=0.50,angle=-0]{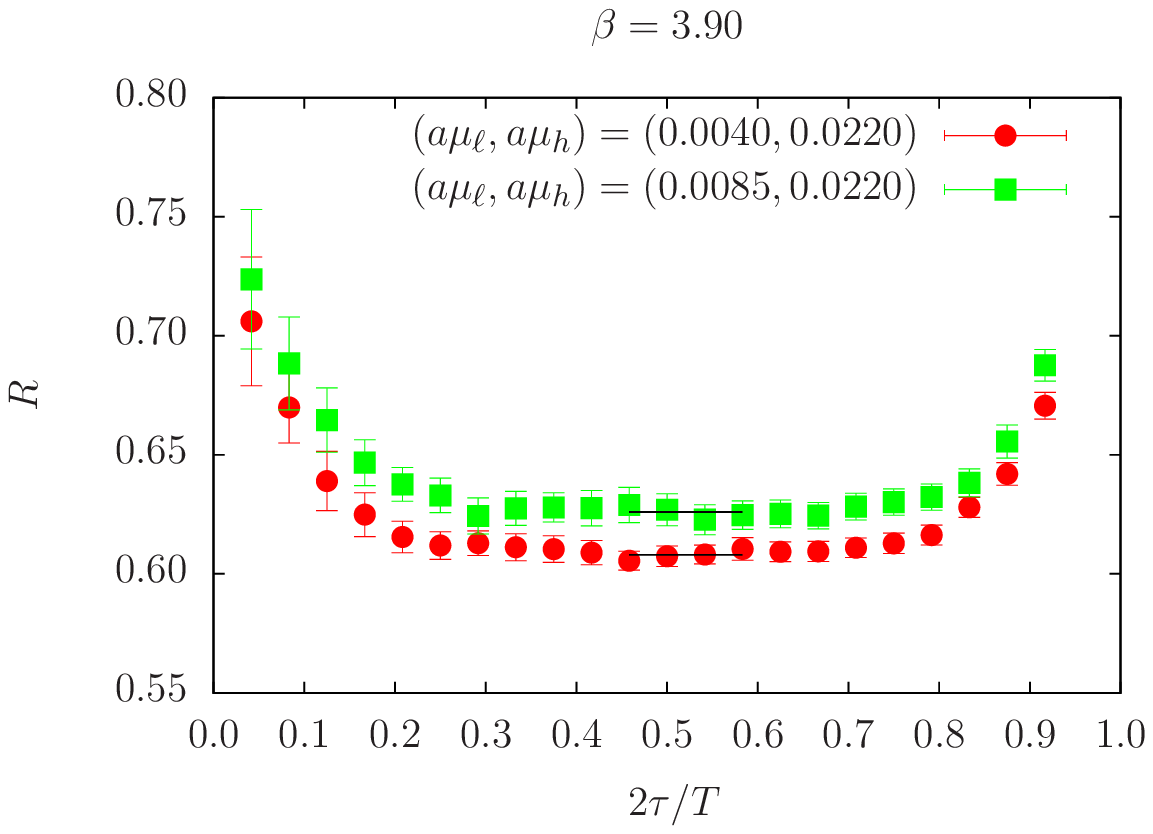}}
\subfigure[]{\includegraphics[scale=0.50,angle=-0]{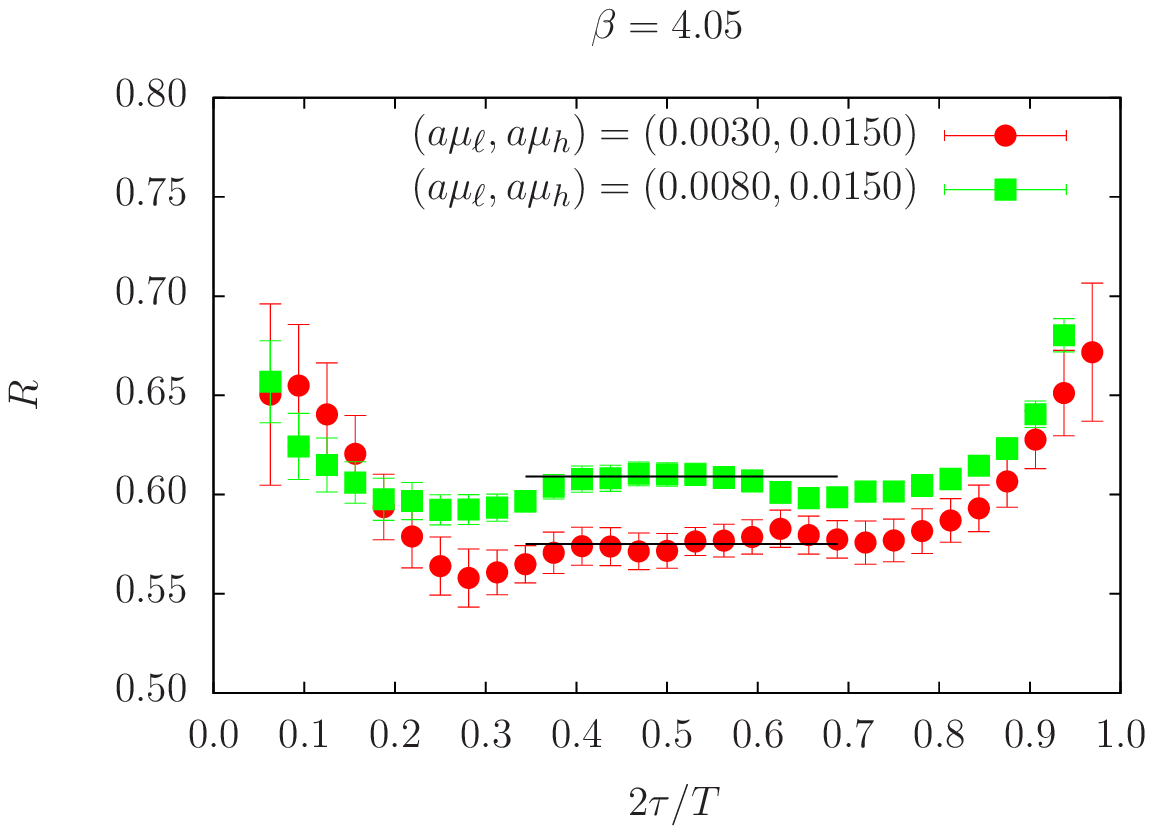}}
\hspace*{2.0cm}\subfigure[]{\includegraphics[scale=0.50,angle=-0]{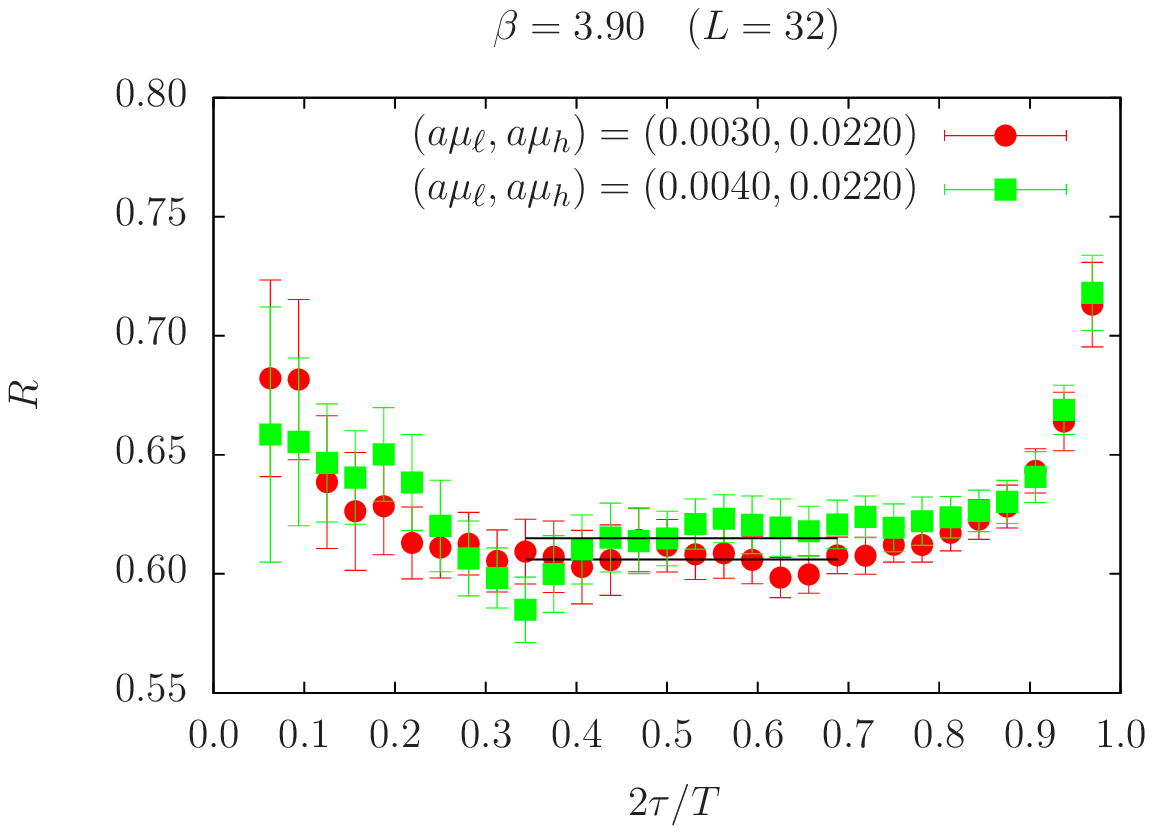}}
\vskip -0.5cm
\begin{center}
\caption{Data and time-plateaux for $R(x_0)$ plotted vs $2\tau/T \equiv 2(x_0-y_0)/T$ 
for two different combinations of light and heavy quark masses: (a) results at $\beta = 3.80$; 
(b) results at $\beta = 3.90$ for the lattice $24^3 \times 48$; (c) results at $\beta = 4.05$; 
(d) results at $\beta = 3.90$ for the lattice $32^3 \times 64$.
}
\label{fig:R}
\end{center}
\end{figure}

\begin{figure}[!ht]
\begin{center}
\includegraphics[scale=0.60,angle=-0]{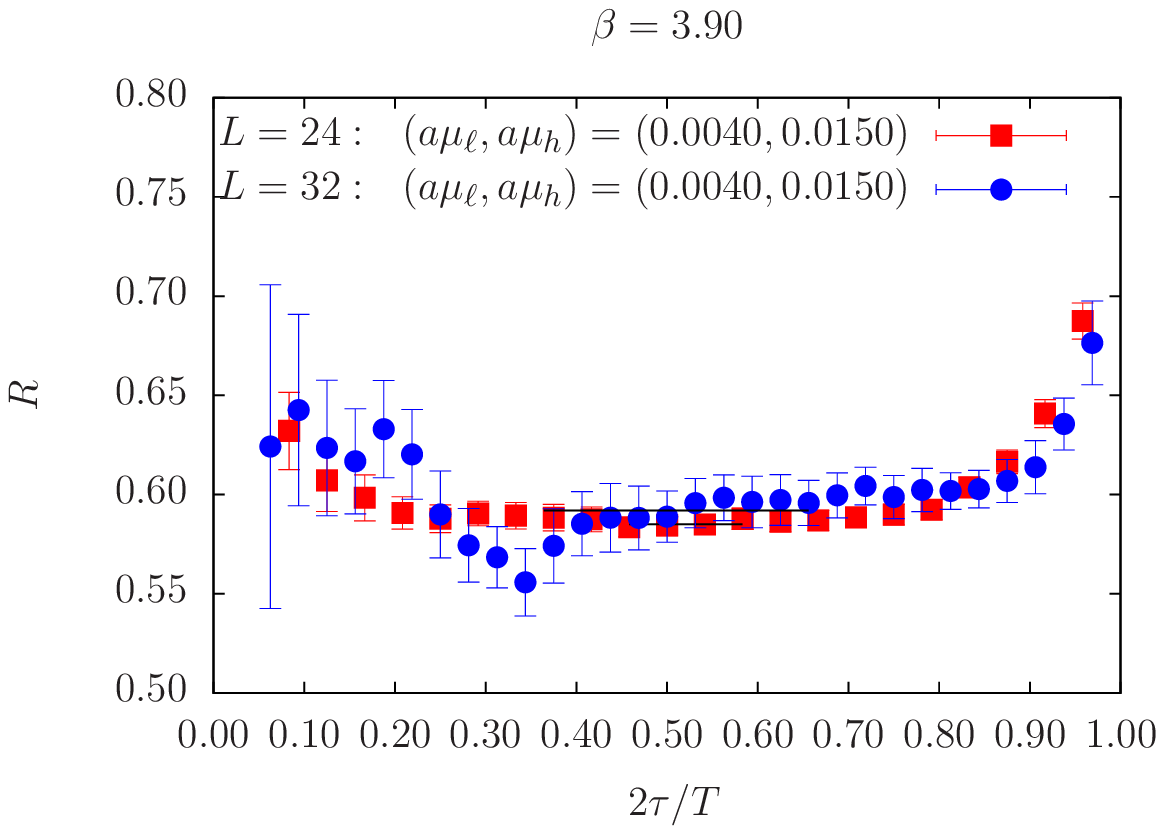}
\end{center}
\vskip -0.5cm
\begin{center}
\caption{Comparison of $R(x_0)$-data plotted vs $2\tau/T \equiv 2(x_0-y_0)/T$ at two different 
lattice volumes for $\beta=3.90$ and the smallest available values of $\mu_\ell$ and $\mu_h$.}
\label{fig:R_FV}
\end{center}
\end{figure}

As one can appreciate from Fig.~\ref{fig:R}, panels (a), (b) and (c), the quality
of our signal for $R(x_0)$ is quite good. Wider plateaux, as well as stronger suppression 
of unwanted vacuum-sector states wrapping around-the-world, are obtained when the
time extension of the lattice is increased by a factor 4/3 going from a $24^3\times 48$ 
to a $32^3 \times 64$ lattice at $\beta=3.9$ (i.e.\ up to $T \sim 5.8$~fm).   
This is illustrated in Fig.~\ref{fig:R} panel (d) where the values of $R(x_0)$ and
its plateau for the larger lattice is displayed. As shown in Fig.~\ref{fig:R_FV}, no 
systematic effect in the plateau value of $R$ is visible at $\beta=3.90$ upon comparing 
results from the lattices $24^3\times 48$ and $32^3\times 64$. 
Similarly by comparing data for $C_2(x_0)$ ($C'_2(x_0)$)
from the two lattices, we could also establish (see Table~\ref{Table:bare390}) the absence of  
significant finite volume effects in the quantities $M_{12}$ and $F_{12}$ ($M_{34}$ and $F_{34}$).
In particular we checked that the marginally significant finite size effect that
one can notice in $aM^{34}$ at $\beta=3.9$ and $a\mu_\ell=0.0040$ has a  
negligible impact on our determination of $B_K^{\rm RGI}$.

\begin{table}
\hspace*{-2.5cm}
\begin{tabular}{|ccccccc|cc|c|}
\hline
\multicolumn{10}{|c|}{ $\beta=3.80$, $a \sim 0.10 $~fm}\\
\multicolumn{10}{|c|}{$L^3 \times T = 24^3\times 48 ~a^4$} \\
\hline
$a\mu_\ell$ &  $a\mu_h$    & $aM^{34}$   & $aM^{12}$ & $aF^{34}$ & $aF^{12}$ & \hspace{-0.15cm}$x_0$-plat.
& $R$
& \hspace{-0.15cm}$x_0$-plat. & $N_{{\rm meas}}$\\
         &   0.0165     & 0.2558(08) & 0.3393(25)  & 0.0894(04) & 0.0883(15) &
                &  0.627(4)      &             &\\
0.0080   &   0.0200     & 0.2731(07) & 0.3532(23)  & 0.0913(04) & 0.0895(15) & \hspace{-0.15cm}[11, 23]
                        &  0.636(3)      &     \hspace{-0.15cm}[11, 14]         & 170\\
         &   0.0250     & 0.2961(07) & 0.3718(21)  & 0.0936(04) & 0.0909(15)  &
              &  0.647(3)      &             & \\
         \hline\hline
         &   0.0165     & 0.2712(04) & 0.3508(16)  & 0.0924(03) & 0.0900(16)   &
              &  0.632(4)      &             &\\
0.0110 & 0.0200 & 0.2877(04) & 0.3644(14)  & 0.0942(03) & 0.0910(16) & \hspace{-0.15cm} [11, 23] &  0.640(4) & 
\hspace{-0.15cm} [11, 14]         & 180\\
         &   0.0250     & 0.3098(04) & 0.3828(12)  & 0.0966(03) & 0.0924(16)    &
             &  0.651(3)      &             &\\
\hline\hline
\end{tabular}
\caption{Pseudoscalar masses, decay constants and the (bare) ratio $R$ at $\beta = 3.80$. The time-plateaux 
``$x_0$-plat.'' in column 7 correspond to the interval $a^{-1}[x_0^{\rm min}, x_0^{\rm max}]$ over which 
the correlators $C_2$ and $C'_2$ have been taken, and in column 9 to the interval 
$a^{-1}[(x_0-y_0)^{\rm min}, (x_0-y_0)^{\rm max}]$ over which the correlator $C_3$ 
has been taken (see sect.~\ref{sec:LATCO}).
$N_{{\rm meas}}$ denotes the number of (well separated) gauge configurations on which the various correlators have been evaluated.}
\label{Table:bare380}
\end{table}

\begin{table}
\hspace*{-2.5cm}
\begin{tabular}{|ccccccc|cc|c|}
\hline
\multicolumn{10}{|c|}{ $\beta=3.90$, $a \sim 0.09 $~fm}\\
\multicolumn{10}{|c|}{$L^3 \times T = 24^3\times 48 ~a^4$} \\
\hline
$a\mu_\ell$ &  $a\mu_h$ & $aM^{34}$ & $aM^{12}$ & $aF^{34}$ & $aF^{12}$ & \hspace{-0.15cm}$x_0$-plat. & $R$
& \hspace{-0.15cm}$x_0$-plat. & $N_{{\rm meas}}$ \\
         &   0.0150     & 0.2060(05) & 0.2639(11)  & 0.0724(03) & 0.0705(09) &
                        &  0.585(5)      &             &                   \\
0.0040   &   0.0220     & 0.2401(05) & 0.2915(11)  & 0.0757(03) & 0.0725(09) & \hspace{-0.15cm} [11, 23] &  0.608(4)      
&   \hspace{-0.15cm}[11, 14]   & 400 \\
         &   0.0270     & 0.2619(05) & 0.3096(11)  & 0.0777(03) & 0.0759(09) &
                        &  0.621(4)      &             &            \\
\hline\hline
         &   0.0150     & 0.2179(08) & 0.2762(16)  & 0.0755(05) & 0.0736(10) &
                         &  0.602(7)      &                    &\\
0.0064   &   0.0220     & 0.2506(07) & 0.3028(15)  & 0.0785(05) & 0.0759(09) &  \hspace{-0.15cm}[11, 23]  &  0.620(6)      
&    \hspace{-0.15cm} [11, 14]        & 200 \\
         &   0.0270     & 0.2717(07) & 0.3204(14)  & 0.0805(04) & 0.0774(09) &
                         &  0.631(6)      &                   &\\
         \hline\hline
         &   0.0150     & 0.2283(07) & 0.2849(17)  & 0.0773(03) & 0.0755(09) &
                         &  0.606(6)      &                   &\\
0.0085   &   0.0220     & 0.2598(07) & 0.3109(15)  & 0.0804(03) & 0.0779(08) & \hspace{-0.15cm} [11, 23]   &  0.626(5)      
& \hspace{-0.15cm}     [11, 14]              & 200 \\
         &   0.0270     & 0.2803(07) & 0.3281(15)  & 0.0823(03) & 0.0794(09) &
                         &  0.637(5)      &                   &\\
                 \hline\hline
         &   0.0150     & 0.2351(07) & 0.2892(14)  & 0.0784(04) & 0.0761(09) &
                         &  0.606(8)      &                   &\\
0.0100   &   0.0220     & 0.2659(07) & 0.3154(12)  & 0.0815(04) & 0.0787(08) & \hspace{-0.15cm} [11, 23]     &  0.625(8)      
& \hspace{-0.15cm}      [11, 14]             & 152\\
         &   0.0270     & 0.2860(06) & 0.3328(12)  & 0.0834(04) & 0.0802(09) &
                         &  0.637(7)      &                   &\\
\hline
\multicolumn{10}{|c|}{$L^3 \times T = 32^3\times 64 ~a^4$} \\
\hline
         &   0.0150     & 0.2041(04) & 0.2644(15)  & 0.0727(03) & 0.0702(11) &
                         &  0.592(5)      &             &                    \\
0.0040   &   0.0220     & 0.2381(04) & 0.2917(17)  & 0.0758(03) & 0.0722(10) & \hspace{-0.15cm} [11, 31]   &  0.615(5)      
& \hspace{-0.15cm}     [11, 22]              & 160
   \\
         &   0.0270     & 0.2599(04) & 0.3096(15)  & 0.0777(03) & 0.0753(09) &
                         &  0.630(5)      &                   &             \\
                 \hline\hline
         &   0.0150     & 0.1982(04) & 0.2558(13)  & 0.0720(03) & 0.0701(07) &
                         &  0.585(4)      &             &                   \\
0.0030   &   0.0220     & 0.2329(04) & 0.2838(11)  & 0.0750(03) & 0.0720(08) & \hspace{-0.15cm} [11, 31]  &  0.606(4)      
& \hspace{-0.15cm}     [11, 22]             & 300
\\
         &   0.0270     & 0.2550(04) & 0.3021(11)  & 0.0770(03) & 0.0745(08) &
                       &  0.619(4)      &                   &             \\

\hline\hline
\end{tabular}
\caption{Same as in Table~\ref{Table:bare380} for $\beta = 3.90$.}
\label{Table:bare390}
\end{table}

\begin{table}
\hspace*{-3cm}
\begin{tabular}{|ccccccc|cc|c|}
\hline
\multicolumn{10}{|c|}{ $\beta=4.05$, $a \sim 0.07 $~fm}\\
\multicolumn{10}{|c|}{$L^3 \times T = 32^3\times 64 ~a^4$} \\
\hline
$a\mu_\ell$ &  $a\mu_h$    & $aM^{34}$   & $aM^{12}$    &  $aF^{34}$ & $aF^{12}$ & \hspace{-0.15cm}$x_0$-plat. 
& $R$    & \hspace{-0.15cm}$x_0$-plat. & $N_{{\rm meas}}$\\
         &   0.0120     & 0.1602(08) & 0.1931(18)  & 0.0564(03) & 0.0558(07) &
                         &  0.560(6)     &             &    \\
0.0030   &   0.0150     & 0.1751(08) & 0.2053(17)  & 0.0578(03) & 0.0566(07) & \hspace{-0.15cm} [14, 31]  &  0.575(6)     
& \hspace{-0.15cm}   [14, 19]  & 190  \\
         &   0.0180     & 0.1889(08) & 0.2169(16)  & 0.0591(03) & 0.0573(07) &
                         &  0.588(6)     &             &                  \\
\hline\hline
         &   0.0120     & 0.1739(06) & 0.2034(11)  & 0.0600(04) & 0.0585(08) &
                         &  0.584(7)     &             &\\
0.0060   &   0.0150     & 0.1877(06) & 0.2153(11)  & 0.0613(03) & 0.0596(08) & \hspace{-0.15cm} [14, 31]  &  0.595(7)     
& \hspace{-0.15cm}   [14, 19]  & 150\\
         &   0.0180     & 0.2007(06) & 0.2266(11)  & 0.0625(03) & 0.0605(08) &
                         &  0.605(6)     &             & \\
         \hline\hline
         &   0.0120     & 0.1840(05) & 0.2127(09)  & 0.0615(04) & 0.0604(12) &
                         &  0.600(6)     &             &\\
0.0080   &   0.0150     & 0.1972(05) & 0.2242(09)  & 0.0627(04) & 0.0616(12) & \hspace{-0.15cm} [14, 31]  &  0.609(5)     
& \hspace{-0.15cm}   [14, 19]  & 220\\
         &   0.0180     & 0.2097(05) & 0.2351(09)  & 0.0638(04) & 0.0626(12) &
                         &  0.618(5)     &             &\\
\hline\hline
\end{tabular}
\caption{Same as in Table~\ref{Table:bare380} for $\beta = 4.05$.}
\label{Table:bare405}
\end{table}

\vskip 0.0cm

\vskip 0.0cm

\subsection{Computing $B_K^{\rm RGI}$ at the physical point}
\label{sec:CBHAT}

In order to arrive at the RGI value of $B_K$ at the physical point
we need to go through the following steps: operator renormalization at each lattice spacing, 
continuum ($a \to 0$) and chiral extrapolation ($\hat\mu_{\ell} \to \hat\mu_{u/d}$) with
$\hat\mu_h$ set to the strange quark mass $\hat\mu_s$ as determined by using the 
experimental value of the kaon mass.

\subsubsection{Computing $Z_{AV+VA}$}
\label{sec:ZAVVA}

Renormalization of $B_K$ requires in particular the calculation of the 
four-fermion operator renormalization constant, $Z_{AV+VA}$, in the mass independent 
scheme. This task 
was carried out using the  RI-MOM approach. 
We followed the general strategy presented in ref.~\cite{Constantinou:2010gr}  
employing the two procedures referred there as M1 (or extrapolation method) 
and M2 (or $p^2$-window method). In this way one
obtains for each $\beta$ two estimates of the renormalization constant at the lattice 
reference scale, which we will denote  
as $Z_{AV+VA}^{{\rm RI{'},M}j}(a^{-2})$, $j=1,2$.
Using the known NLO anomalous dimension of the four fermion operator in the RI-MOM
scheme, the RGI quantities $Z_{AV+VA}^{{\rm RGI,M}j}$ are evaluated. We note that the
cutoff effects on the lattice estimators of $Z_{AV+VA}$ are significantly reduced by subtracting  
their 1-loop perturbative expression up to O($a^2 g_0^2$),  
computed in~\cite{HP4ferm}. Further information on the
computation of the four-fermion operator renormalization constant, including a numerical 
check on the absence of wrong chirality  
mixing, is given in Appendix~A. 

In the following we will quote results obtained by employing the estimates of $Z_A$
and $Z_{AV+VA}^{{\rm RGI}}$ from the procedure M2, as well as the very precise 
determination of $Z_V$ obtained in ref.~\cite{Constantinou:2010gr}. 
The latter is determined by comparing (matrix elements of) the local current to the exactly conserved one. 
We immediately notice that upon using the procedure M1 for the 
evaluation of $Z_A$ and $Z_{AV+VA}^{{\rm RGI}}$ 
very similar results for $B_K^{\rm RGI}$ are obtained, with differences typically
as small as 0.002 (that we include in the systematic error)~\footnote{The  
 consistency 
of the continuum limit results for the unphysical quantity 
$B_K^{\rm RGI}(\hat\mu_\ell^*, \hat\mu_h^*)$ 
evaluated with renormalization constants from the procedure M1 or M2 
was  noted in sect.~\ref{sec:SCACH}.}. 
The  values of the renormalization constants entering the
present computation of $B_K^{\rm RGI}$ can be found in Table~\ref{tab:A-RCval}.

\subsubsection{Continuum and chiral extrapolations}
\label{sec:CCE}

Continuum and chiral ($\mu_\ell \to \mu_{u/d}$) extrapolations are carried out 
simultaneously exploiting the SU(2)-$\chi$PT based fit ansatz of ref.~\cite{Allton:2008pn}  
which, for our computation with $\mu_{sea}=\mu_\ell$, reads~\footnote{With respect 
to the formulae of ref.~\cite{Allton:2008pn} we replace $\Lambda_{\chi}$ with $4\pi f_0$
and reabsorbe the effect of this change in a redefinition of $b(\hat\mu_h)$.
We use $f_0 \simeq 121$~MeV as found in~\cite{Baron:2009wt}.}
\begin{equation} \label{BCHIR2}
B_{K,{\rm lat}}^{\rm RGI}(\hat\mu_\ell,\hat\mu_s)= B_{K,\chi}^{\rm RGI}(\hat\mu_s)
\left[1 +b(\hat\mu_s)\dfrac{2\hat B_0\hat\mu_{\ell}}{f_0^2}-
\dfrac{2\hat B_0\hat\mu_\ell}{32\pi^2 f_0^2} 
\log \dfrac{2\hat B_0\hat\mu_\ell}{16\pi^2 f_0^2} \right]+
a^2 f_0^2 D_B(\hat\mu_s)\, ,
\end{equation}
where $B_{K,{\rm lat}}^{\rm RGI}(\hat\mu_\ell,\hat\mu_s)$ denotes the RGI
lattice estimates of the ``bag parameter'' at renormalized $\ell$-quark mass
$\hat\mu_\ell$ and strange quark mass $\hat\mu_s$. Note that 
all dimensionful quantities are expressed in units of $f_0$. The
dimensionless fit parameters $B_{K,\chi}^{\rm RGI}(\hat\mu_s)$, $b(\hat\mu_s)$
and $D_B(\hat\mu_s)$ are functions of $\hat\mu_s$. The first of them
represents the RGI ``bag parameter'' at $\hat\mu_s$  
 in the SU(2)-$\chi$PT  
limit ($\hat\mu_{\ell} \to 0$).

Several comments are in order here. First, as shortly mentioned above, the renormalized 
(in $\overline {MS}$ at 2~GeV) quantities 
$\hat B_0$ and $\hat \mu_{u/d}$ and the chiral limit pion decay constant $f_0$
\begin{equation} \label{B0mdfpi}
\hat B_0 = 2.84(11)~{\rm GeV}\, , \qquad 
\hat \mu_{u/d} = 3.5(1)~{\rm MeV}\, , \qquad
f_0 = 121.0(1)~{\rm MeV}\, , 
\end{equation}
as well as estimates of $a$ at $\beta=3.8,3.9,4.05$,
\begin{equation} \label{a-vals}
af_0|_{\beta=3.8} = 0.0604(16) \, , \qquad
af_0|_{\beta=3.9} = 0.0514(08) \, , \qquad
af_0|_{\beta=4.05} = 0.0409(07) \, , 
\end{equation}
are obtained by an 
analysis of the data for the mass and the decay constant of the $\ell\ell$ (charged) pseudoscalar 
meson following the lines of ref.~\cite{Baron:2009wt} and using the results
on the quark mass renormalization constant $Z_\mu = Z_P^{-1}$ from~\cite{Constantinou:2010gr}.
Here we just recall that in this analysis based on SU(2)-$\chi$PT at NLO we
set the physical scale using the experimental value of $f_\pi$ and take
into account finite-size effects and O($a^2$) artifacts. 
As we rely on NLO SU(2)-$\chi$PT formulae, the $\ell\ell$ pseudoscalar meson data are taken only
from a $\hat\mu_\ell$ region (see Tables 1, 2 and 3) where the impact of possible NNLO corrections has been 
checked (via the methods of ref.~\cite{Baron:2009wt}) to be negligible.

The renormalized strange quark mass $\hat\mu_s$ is extracted by analysing our data 
on the $h\ell$ pseudoscalar meson mass $M_{34}$ at $\beta=4.05$, $3.9$ and $3.8$ 
with the help of the SU(2) $\chi$PT based ansatz~\cite{Allton:2008pn}
\begin{equation} \label{MK2FIT} 
f_0^{-2}M_{34}^2(\hat\mu_\ell,\hat\mu_h) = C_M(\hat\mu_h)
\left[ 1 + c(\hat\mu_h)\dfrac{2\hat B_0\hat\mu_{\ell}}{f_0^2} \right]+
a^2 f_0^2 D_M(\hat\mu_h)\, ,
\end{equation} 
for three reference values of $\hat\mu_h$ in the strange quark mass region (from 75 to 105 MeV). 
In the fit ansatz~(\ref{MK2FIT}), with all dimensionful quantities expressed in units of $f_0$,
we have included a term, with coefficient $D_M(\hat\mu_h)$, to parameterize the leading O($a^2$) 
artifacts. Finite size effects on the raw lattice data for $M_{34}^2$, although not shown in 
eq.~(\ref{MK2FIT}), were taken into account in our actual fits. These effects,
which are checked at $\beta=3.9$, $a\mu_\ell=0.0040$ to be $\leq 1\%$
and only marginally significant within errors (see Table~\ref{Table:bare390}), have been estimated 
by using resummed $\chi$-PT formulae~\cite{CDH} and subtracted out from the lattice data.
The coefficients $C_M(\hat\mu_h)$ and $D_M(\hat\mu_h)$ grow almost linearly with $\hat\mu_h$
in the aforementioned range, 
in line with the chiral behaviour expected at finite lattice spacing, namely $M_{34}^2 \sim 
(\hat\mu_\ell + \hat\mu_h)$ as $(\hat\mu_\ell + \hat\mu_h) \to 0$~\cite{Frezzotti:2005gi}. 

Since at this stage the value of $\hat\mu_{u/d}$ is already known, 
once the parameters $C_M(\hat \mu_h)$, $c(\hat\mu_h)$ and 
$D_M(\hat\mu_h)$ have been determined by the fit of the 
$\hat\mu_\ell$-dependence based on eq.~(\ref{MK2FIT}), 
one immediately yields continuum limit estimates of $M_{34}^2(\hat\mu_{u/d},\hat\mu_h)$ 
for the three chosen reference values of $\hat\mu_h$. 
These estimates exhibit a smooth dependence on $\hat\mu_h$ and can be interpolated 
linearly to the experimental value of $M_K^2= (495 {\rm MeV})^2$, thereby providing 
a well controlled result for $\hat\mu_s$, for which we find 
\begin{equation} \label{mus-res}
 \hat\mu_s \equiv \mu_s^{(\overline{\rm MS},~2~{\rm GeV})} \; = \; 92(5) ~{\rm MeV} \, .
\end{equation}
More details on this analysis will be given in a forthcoming publication~\cite{ETMC:qmass-prep}.

\vskip 0.0cm
\begin{figure}[!ht]
\begin{center}
\subfigure[]{\includegraphics[scale=0.50,angle=-0]{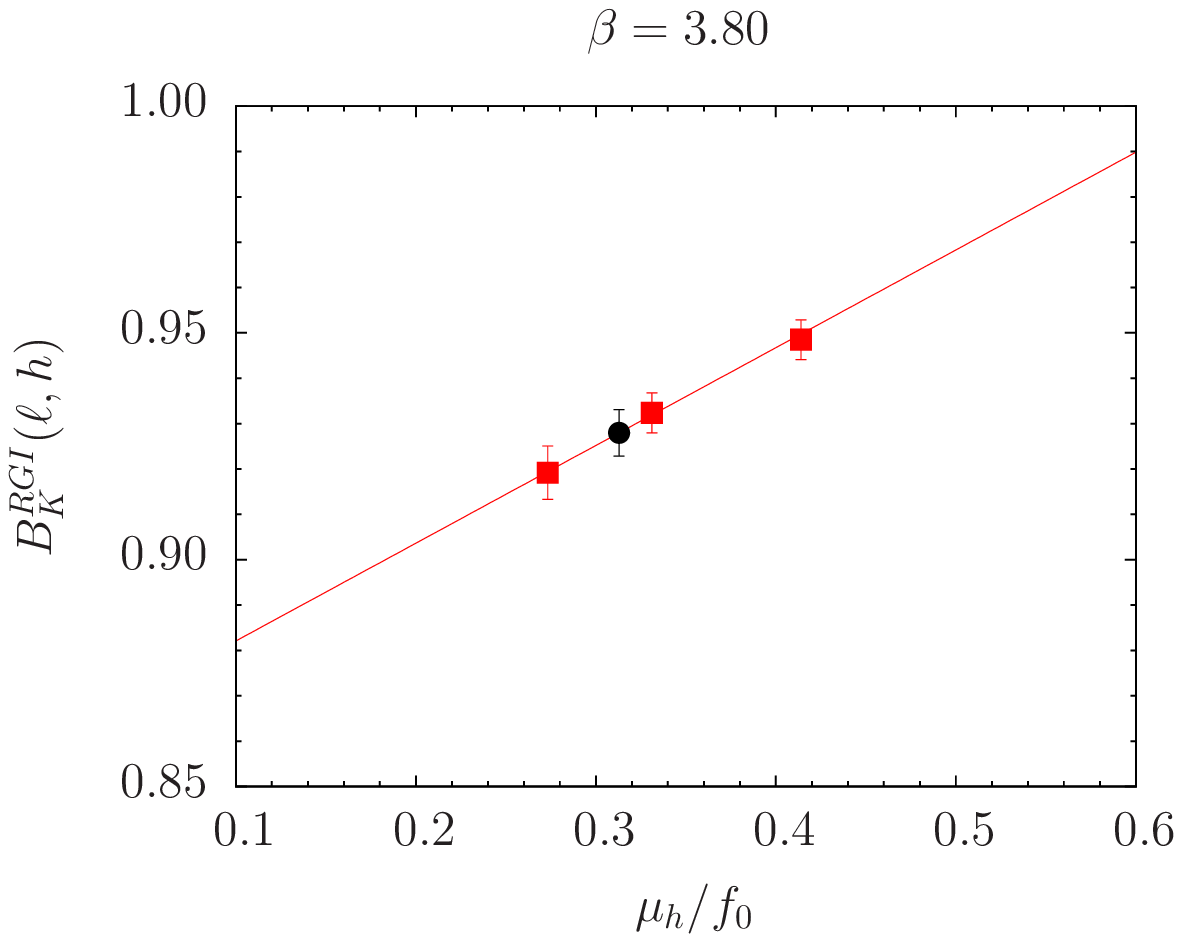}}
\subfigure[]{\includegraphics[scale=0.50,angle=-0]{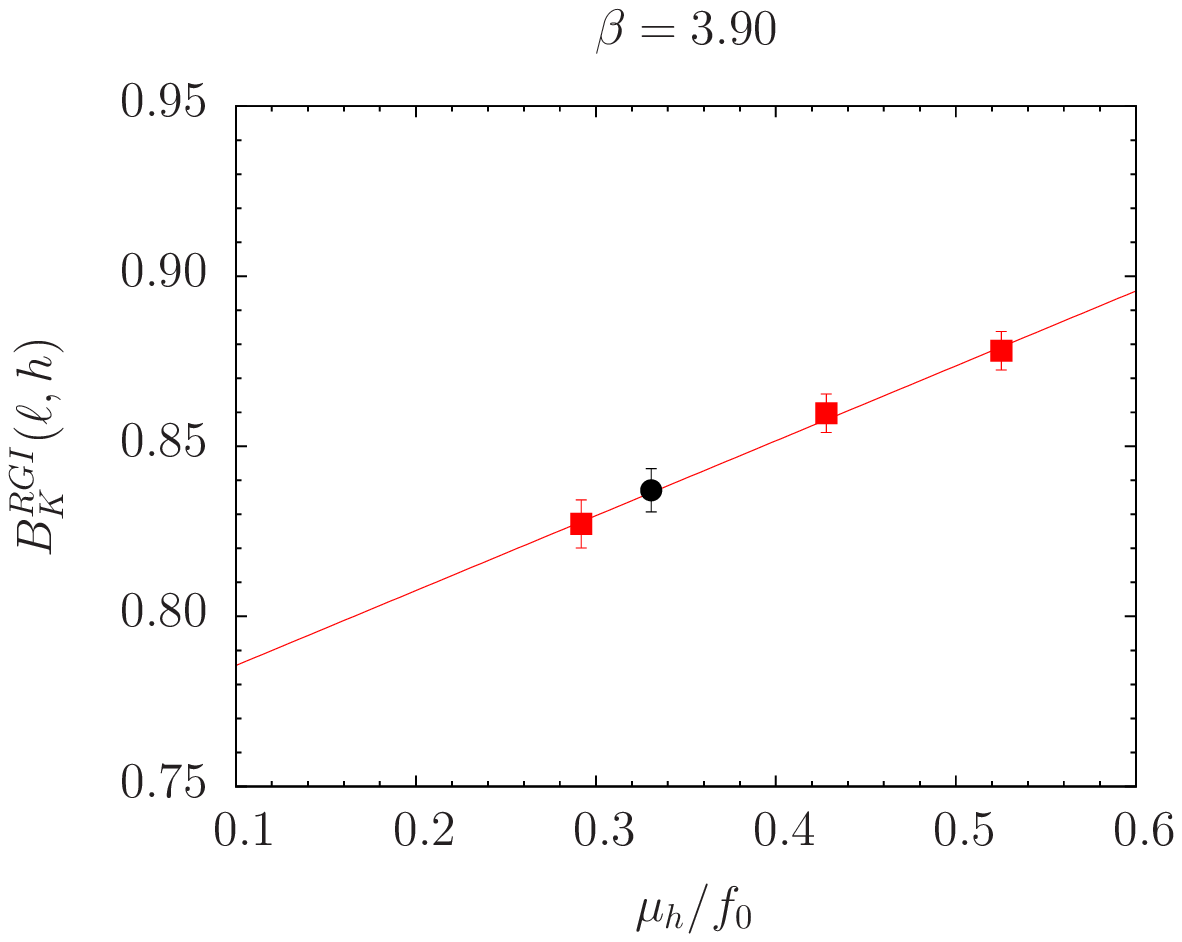}}
\\
\subfigure[]{\includegraphics[scale=0.50,angle=-0]{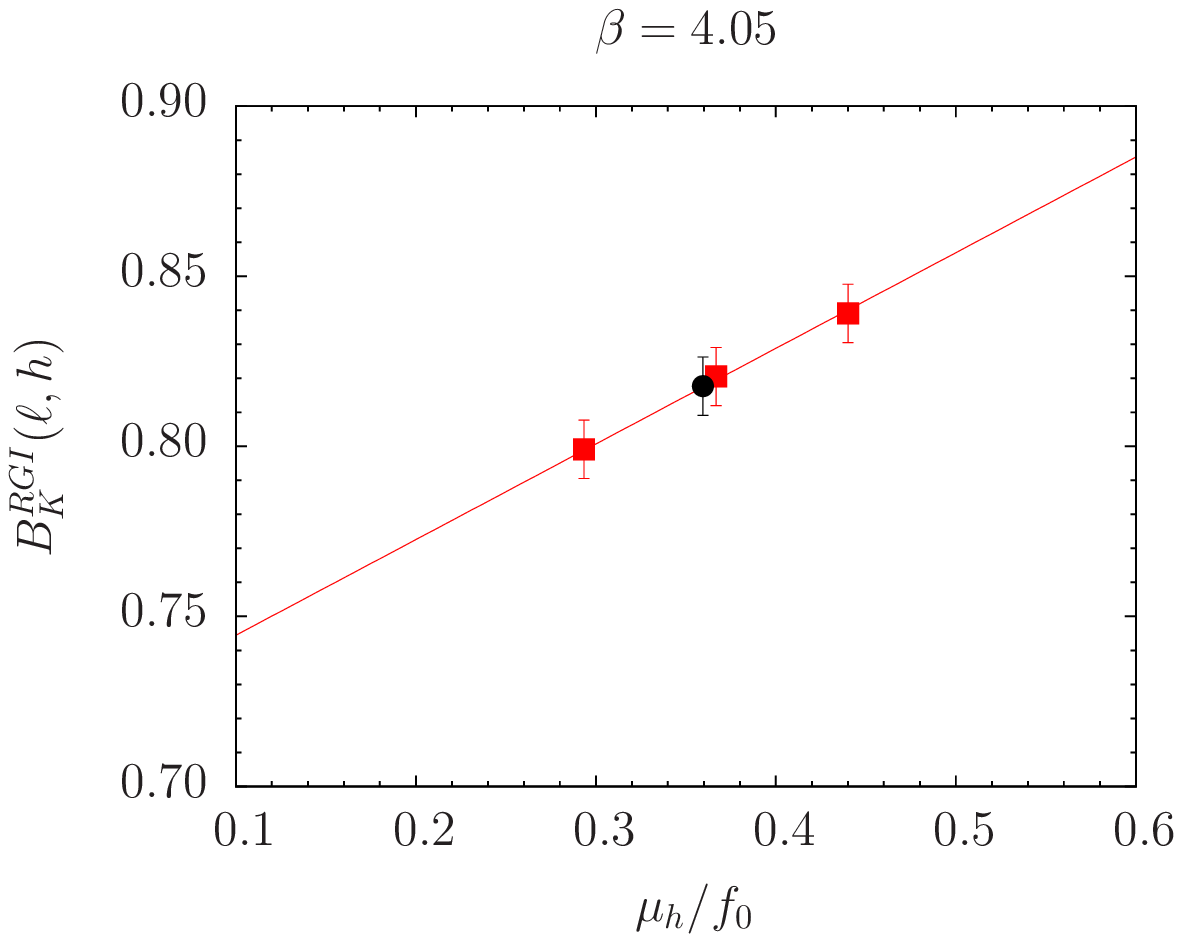}}
\end{center}
\vskip -0.5cm 
\begin{center}
\caption{Linear interpolation of $B_K^{\rm RGI}(\hat\mu_\ell,\hat\mu_h)$ 
 to the physical strange quark mass point $\hat\mu_h \to\hat\mu_s = 92$~MeV 
at $(\beta=3.8,~a\mu_\ell=0.0080)$, panel (a), $(\beta=3.9,~a\mu_\ell=0.0040)$, 
panel (b), and $(\beta=4.05,~a\mu_\ell=0.0030)$, panel (c).
}
\label{fig:BKM2M2_interp}
\end{center}
\end{figure}

\begin{figure}[!ht]
\begin{center}
\includegraphics[scale=0.70,angle=-0]{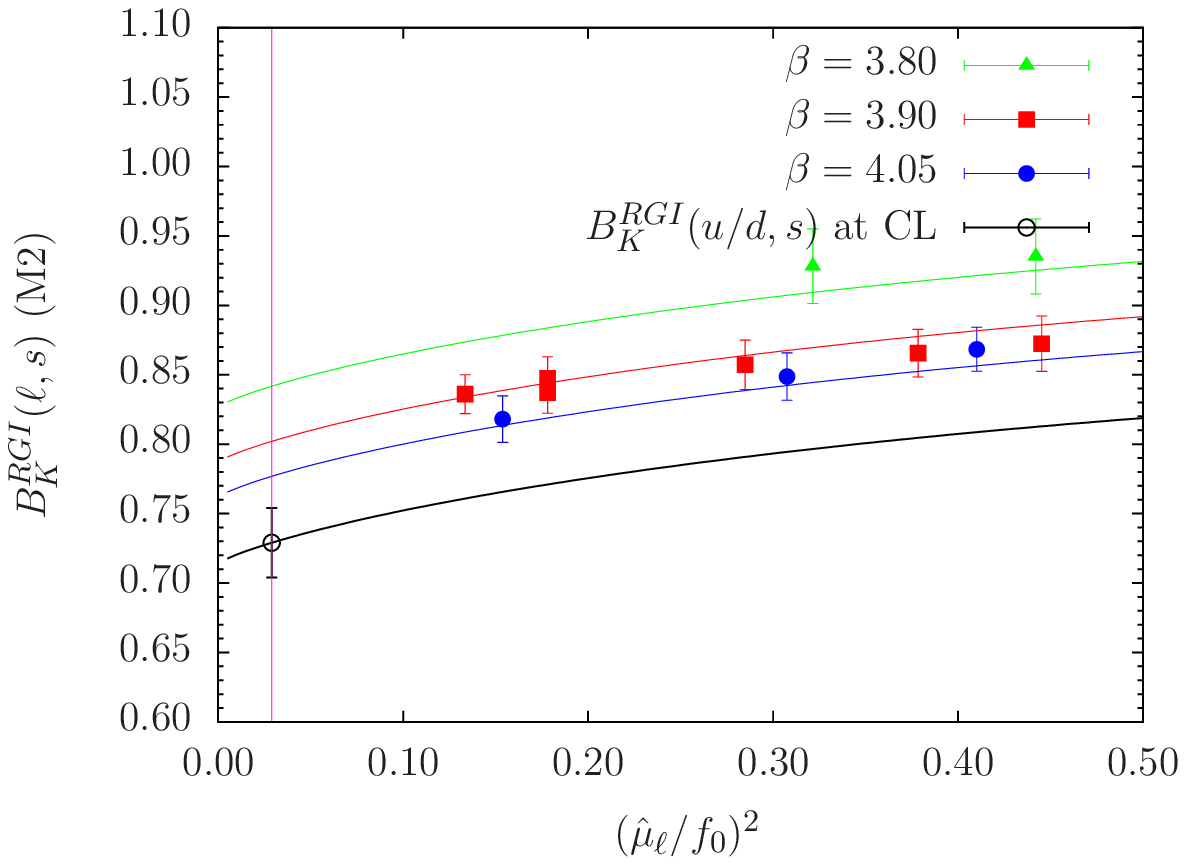}  
\end{center}
\vskip -0.5cm
\begin{center}
\caption{Combined chiral and continuum fits according to the ansatz~(\ref{BCHIR2}). The label M2  
refers to the specific RI-MOM procedure adopted for the computation
of $Z_{VA+AV}^{\rm RGI}$ and $Z_A$. 
}
\label{fig:BKM2M2_cochfit}
\end{center}
\end{figure}

Also our lattice estimators for $B_{K,{\rm lat}}^{\rm RGI}(\hat\mu_\ell,\hat\mu_h)$
turn out to have a weak dependence on $\hat\mu_h$. It is thus straightforward
to obtain by linear interpolation in $\hat\mu_h$ their values at the point 
$\hat\mu_h = \hat\mu_s$. By exploiting the available determinations of $a(\beta)$
in physical units from the analysis of the pion sector data, the interpolation of
$B_{K,{\rm lat}}^{\rm RGI}(\hat\mu_\ell,\hat\mu_h)$ to
the $s$-quark mass point can be performed separately at each value of $\beta$
and $a\hat\mu_\ell$. An example of such interpolations is shown in Fig.~\ref{fig:BKM2M2_interp}.
In this way we obtain the quantity $B_{K,{\rm lat}}^{\rm RGI}(\hat\mu_\ell,\hat\mu_s)$
for the three lattice spacing and the $a\hat\mu_\ell$-values corresponding to the bare
parameters in Tables~\ref{Table:bare380}, \ref{Table:bare390} and~\ref{Table:bare405}. 
This set of numbers can be fit 
to the formula~(\ref{BCHIR2}) where the quantities $B_{K,\chi}^{\rm RGI}(\hat\mu_s)$, 
$b(\hat\mu_s)$ and $D_B(\hat\mu_s)$ are the free parameters to be determined. 
As it clearly appears from Fig.~\ref{fig:BKM2M2_cochfit} 
the quality of the resulting fit is very good.

\vskip 0.0cm

\subsubsection{Final result and error budget}
\label{sec:FREB}

The value of $B_K^{\rm RGI}$ at the physical point is finally 
determined by evaluating eq.~(\ref{BCHIR2}) in the continuum limit  
at the best-fit values of $B_{K,\chi}^{\rm RGI}(\hat\mu_s)$ and $b(\hat\mu_s)$ 
and setting $\mu_\ell = \mu_{u/d}$. In this way we arrive at the result
\begin{equation} \label{BK-f0M2M2}
B_K^{\rm RGI} = 0.729(25) \, .
\end{equation}
The error quoted in parentheses in eq.~(\ref{BK-f0M2M2}) is  
of statistical origin. 
It comes from our fitting procedure and reflects the 
statistical errors on the raw data for the bare matrix elements and the associated 
renormalization constants. Its estimate
has been performed via a standard bootstrap analysis, thereby taking properly into
account all possible cross-correlations. 
As one checks in Tables~\ref{Table:bare380}--\ref{Table:bare405},
at fixed quark masses and lattice spacing the typical statistical error on the bare 
lattice estimator of $B_K$ is around 1\%. Inspection of Table~\ref{tab:A-RCval} 
shows that relative error coming from the $B_K$-renormalization factor 
$Z_{VA+AV}^{\rm RGI}/(Z_AZ_V)$ is about 2.0\%. The extrapolation to the  
continuum limit and the physical pion mass, plus interpolation to the kaon point,
finally leads to the error quoted in eq.~(\ref{BK-f0M2M2}).

In order to try to have an idea of the size of the systematic error 
other analyses have been performed besides the one discussed above.

i)  First of all, the whole analysis was repeated on
the same set of raw data but choosing $a/r_0$ extrapolated to the chiral limit
~\cite{Baron:2009wt} (with $r_0$ the Sommer scale),
instead of $af_0$, as the scaling variable used to build dimensionless
quantities suited for continuum extrapolation. This change turned out to produce 
an increase of the value of $B_K^{\rm RGI}$ of a few permille.

ii) Secondly we replaced $2\hat B_0\hat \mu_\ell$ by the squared $\ell\ell$-meson
mass $(M_{\ell\ell}^{\rm tm})^2$ in extrapolating to the physical 
pion mass point and we determined the kaon point by interpolating the quantity 
$(M_{hh}^{\rm tm})^2$ to the physical value of $2M_K^2 - M_\pi^2$.
As before $B_K^{\rm RGI}$ increases by a few permille. 
Moreover we have tried a fit function which also includes higher order analytical terms
added to the SU(2)$-\chi$PT formula. We observe a shift equal to 0.007  in the central
value of $B_{\rm{K}}^{\rm{RGI}}$.

iii) We also tried to perform the extrapolation to the physical pion mass point 
using a first or second order polynomial in $\mu_\ell$ in alternative
to the $SU(2)$$\chi$PT-based ansatz~(\ref{BCHIR2}). We obtained good
quality fits where the central value of $B_K^{\rm RGI}$ gets shifted by +0.021 or +0.001, 
respectively. 
Completely analogous results are obtained by employing a first or second
order polynomial in $(M_{\ell\ell}^{\rm tm})^2$.

iv) Finally all these alternative analyses were repeated  
by using the M1 method for the 
RI-MOM renormalisation in place of M2.  In
all cases this produced tiny changes of $B_K^{\rm RGI}$ (never larger
than $\pm 0.004$). For instance,   
for the main analysis discussed
in some detail in the present section, a decrease of $B_K^{\rm RGI}$
by 0.002 was observed. Further information on the analyses for the estimate
of the systematic error is given in Appendix~B. 

In this way we arrive at the systematic error budget for $B_K^{\rm RGI}$
we anticipated in the Introduction, which is detailed below. We get 
\begin{itemize}
\item 0.004 from the RI-MOM renormalization, evaluated by combining the uncertainty 
       due to the spread coming from methods M1 and M2 and the estimated error due the truncation 
       to NLO of the perturbative series in the conversion from the RI-MOM scheme to RGI.
\item 0.009 from the uncertainty in controlling/removing cutoff effects.
       In this figure we combine the neglected O($a^4$) lattice artifacts which we estimate to be
       $\sim  1\%$~\footnote{The estimate of $\sim  1\%$
       as the natural size of the O($a^4$) corrections 
       follows from the fact that we observe in our $B_K^{\rm RGI}$ lattice estimators 
       O($a^2$) cutoff effects at a level of $\sim 10\%$.}
       and the effect related to the choice of the scaling variable used in the analysis ($af_0$ or $a/r_0$). 
\item  0.014 from the systematic uncertainty in the extrapolation
       to the pion mass point, where we add in quadrature 0.004 from varying the choice of
       the extrapolation variable ($\mu_\ell$ or $(M_{\ell\ell}^{\rm tm})^2$),
       0.011 from the difference between an $SU(2)$$\chi$PT-based fit
       and a simple polynomial ansatz (we take the average of the spread 
       resulting from the polynomial fits discussed above)  and a shift equal to  0.007 
       due to a possible inclusion of higher order analytical terms in the SU(2)$\chi$PT fit 
       formula.  
        . 
\end{itemize}  
Summing in quadrature all these small systematic uncertainties leads to
a total systematic error estimate of $\pm 0.017$. This number was added
in quadrature to the statistical error in eq.~(\ref{BK-f0M2M2}),
yielding the final result quoted in eq.~(\ref{BK-fin}). 

\section{Conclusions}
\label{sec:concl}

In this paper we have computed from the $N_f=2$ tm-LQCD simulation data 
produced by the ETM Collaboration the value of the strong interaction parameter, 
$\hat B_{\rm K}$, which controls the $K^0-\bar K^0$ oscillations. To this end we 
have exploited the partially quenched set up proposed in ref.~\cite{Frezzotti:2004wz} 
which,  
at the  price of introducing O($a^2$) unitarity violations, ensures O($a$) 
improvement and absence of wrong chirality mixing effects~\cite{Bochicchio:1985xa}. 
Renormalization is carried out non-perturbatively
in the RI-MOM scheme~\cite{renorm_mom:paper1,Constantinou:2010gr}.

Using data at three lattice spacings and a number of pseudoscalar masses in the 
interval 280~MeV $< m_{\rm PS} < 550$~MeV, we get in the continuum limit of  
$N_f=2$
QCD and  at the physical value of the pion and kaon mass the value of the RGI
``bag parameter'' reported in eq.~(\ref{BK-fin}), namely $B_K^{\rm RGI} = 0.729 \pm 0.030$.

Our determination is quite accurate, with a $\sim 4$\% total error (with statistical and 
total systematic errors added in quadrature), and agrees rather well with other 
existing values. 
In particular, from the comparison of our result for $B_{\rm{K}}$ with the results from $N_f =
2 + 1$  dynamical simulations (see Fig.~\ref{fig:figcomp}), 
it may be inferred that the quenching of the strange quark leads to an error which is rather small i.e. 
smaller, at present, than  other
systematic uncertainties affecting current $B_{\rm{K}}$ estimates.
If taken at face value, our result confirms the tension between 
the lattice determination (which do not seem to depend significantly on the number of 
dynamical quark flavours) and the preferred value of recent 
phenomenological, Standard Model based analyses (see 
refs.~\cite{Buras:2008nn,VandeWater:2009uc,Bona:2009zz,Buras:2010pz,Lubicz:2010nx}). 
However some theoretical concerns remain about the way of estimating the 
long distance contributions in the relation between
the strong interaction parameter $\hat B_{\rm K}$ and the experimentally
well measured quantity $\epsilon_K$ that parameterizes
the (indirect) CP violation in the $K^0-\bar K^0$ system.

In order to get a substantially more accurate determination of $B_K^{\rm RGI}$ 
we would need to  
increase the statistics and work at even finer 
lattice spacings with the $u/d$-quark mass closer to the physical  point. A further 
possibly important improvement is the use of data from more realistic simulations 
where dynamical strange and charm sea quarks are included. 
Actually the ETM Collaboration is already 
moving ahead in these directions~\cite{Baron:2010bv} and work aiming at 
more precise lattice determinations of the kaon ``bag parameter'' is in progress.

\vspace{0.5cm}
{\bf{Acknowledgments - }} We wish to thank all the other members 
of ETMC for their interest in this work and a most enjoyable and fruitful collaboration. 
This work was supported by the European Union Sixth
Framework Programme under contract MTRN-CT-2006-035482, FlaviaNet.  
V.~G. thanks the MICINN (Spain) for partial support under grant FPA2008-03373
and the Generalitat Valenciana (Spain) for partial support under grant
GVPROMETEO2009-128.
F.~M. acknowledges financial support from projects
FPA2007-66665, 2009SGR502 and from the Spanish Consolider-Ingenio 2010 Programme
CPAN (CSD2007-00042).
M.~P. acknowledges financial support by a Marie Curie European
Reintegration Grant
of the 7th European Community Framework Programme under contract number
PERG05-GA-2009-249309. A.S. acknowledges financial support from the Research 
Promotion Foundation of Cyprus (Proposal Nr: ENI$\Sigma$X$/0506/17$).

\vspace{1.2cm}

\renewcommand{\thesection}{A}
\section*{Appendix A -- The RI-MOM computation of $Z_{VA+AV}$}
\label{sec:APP-A}

Here we discuss the non-perturbative computation of the renormalization
constant (RC) of the operator $Q_{VV+AA}$ (see eq.~(\ref{Q_val})), which,
for the reasons explained in sect.~\ref{sec:LATOP}, we call $Z_{VA+AV}$.  
It is first evaluated in the RI'-MOM scheme~\footnote{The prime in the label ${\rm RI'}$ is to 
remind the specific  
defining condition adopted for the quark field RC $Z_q$, which, 
as convenient in lattice studies, is taken here to coincide with the quark propagator 
form
factor $\Sigma_1$. For the latter we adopt the definition   
of eq.~(33) of ref.~\cite{Constantinou:2010gr}.}, 
at some appropriate scale $\mu_0$ lying in the domain of applicability of (RG-improved) perturbation theory, 
following the strategy detailed in ref.~\cite{Constantinou:2010gr}. Then $Z_{VA+AV}^{\rm RI'}$
is converted into the (scheme- and scale-independent) quantity $Z_{VA+AV}^{\rm RGI}$ according to
the formula 
\begin{equation}
Z_{VA+AV}^{\rm RGI} = (\alpha(\mu_0))^{-\frac{\gamma_0}{2\beta_0}}
\exp \Big{\{} \int_0^{\alpha(\mu_0)} d\alpha' \Big{[}\dfrac{\gamma_{_{VA+AV}}(\alpha')}{\beta(\alpha')}+
\dfrac{\gamma_0}{2\beta_0\alpha'}\Big{]}\Big{\}}
Z_{VA+AV}^{\rm RI'}(\mu_0^2) \, ,
\label{Z4F-RGI} \end{equation}
where 
\begin{eqnarray}
& \beta(\alpha) = -\dfrac{2\beta_0}{4\pi} \alpha^2 + {\rm O}(\alpha^3) \, , \qquad
\beta_0=11-\dfrac{2N_f}{3}\, , \nonumber \\ 
& \gamma_{VA+AV}(\alpha) = \dfrac{\gamma_0}{4\pi} \alpha + {\rm O}(\alpha^2) \, , \qquad  \gamma_0=4 \, ,
\label{BG}
\end{eqnarray}
which is here evaluated with $N_f=2$ and to NLO accuracy~\cite{NLOevol4F,Buras:2000if}.
RGI quantities are normalized following the conventions of ref.~\cite{Buras:1998raa}.
In this appendix we quote results for $Z_{VA+AV}^{\rm RGI}$ but, since the conversion
from RI'-MOM to RGI is a straigtforward step, most of our discussion will actually
concern the evaluation of $Z_{VA+AV}^{\rm RI'}(\mu_0^2)$ at the three lattice spacings 
of interest for the present work. 
The error on $B_K^{\rm{RGI}}$ associated to the NLO perturbative conversion from RI'-MOM to RGI 
is estimated to be about 0.002 and has been taken into account in the error budget discussion 
of sect.~\ref{sec:FREB}.  
This estimate  is obtained assuming that the neglected NNLO and
higher order relative corrections to the conversion factor are as large as the square of the 
NLO relative correction, which turns out to be $\simeq 5\%$. 

For the purpose of determining $Z_{VA+AV}^{\rm RI'}$, 
besides the $q_f$-quark propagator~\footnote{The equations below are written in the
so-called physical quark basis   
with the lattice valence quark action  given by eq.~(\ref{OSvalact}).}
\begin{equation} 
S_{q_f}(p) = a^4 \sum_x e^{-ipx} \langle q_f(x) \bar q_f(0)  \rangle \, ,
\label{qprop} \end{equation}
we compute the relevant Green function is the momentum space correlator
\begin{eqnarray}
& G_{VV+AA}(p,p,p,p)_{\alpha\beta\gamma\delta}^{abcd} =  
 a^{16} \sum_{x_1,x_2,x_3,x_4} \; e^{-ip(x_1-x_2+x_3-x_4)} \cdot
\nonumber \\
& \cdot \,\langle [q_1(x_1)]_\alpha^a  \, [\bar q_2(x_2)]_\beta^b \,  Q_{VV+AA}(0)  \,
[q_3(x_3)]_\gamma^c \, [\bar q_4(x_4)]_\delta^d \rangle \, . 
\label{5pointCF} \end{eqnarray}  
The where lower-case Greek (Latin) symbols denote uncontracted spin (colour) indices.
In terms of the corresponding amputated Green function~\footnote{We explicitly display in the l.h.s.\ of the equation 
the dependence on quark mass parameters that is implicit in the quantities appearing in the r.h.s.}
\begin{eqnarray}
& \Lambda_{VV+AA}(p,p,p,p;\mu_{val},\mu_{sea})_{\alpha\beta\gamma\delta}^{abcd} \; = \;  
[S_{q,1}(p)^{-1}]_{\alpha\alpha'}^{aa'} \, 
[S_{q,3}(p)^{-1}]_{\gamma\gamma'}^{cc'}  \cdot 
\nonumber \\
& \cdot \,G_{VV+AA}(p,p,p,p)_{\alpha'\beta'\gamma'\delta'}^{a'b'c'd'} 
\, [S_{q,2}(p)^{-1}]_{\beta'\beta}^{b'b} \,
[S_{q,4}(p)^{-1}]_{\delta'\delta}^{d'd} \, ,
\label{VVAAvertex} \end{eqnarray}
the RI'-MOM RC is obtained by imposing in the chiral limit ($\mu_{val}=\mu_{sea}=0$)
the renormalization condition 
\begin{equation}
[Z_{VA+AV}^{\rm RI'} \, Z_q^{-2} \, 
D_{11}](p;\mu_{val},\mu_{sea}) = 1 \, ,  
\label{VVAArencon} \end{equation}
where $Z_q$ denotes the (flavour independent) quark field RC and 
the $VV+AA$-vertex $D_{11}^{\rm lat}$ is given by
\begin{equation} 
D_{11}(p;\mu_{val},\mu_{sea}) = 
  \Lambda_{VV+AA}(p,p,p,p;\mu_{val},\mu_{sea})_{\alpha\beta\alpha'\beta'}^{aabb} 
  {\cal P}_{\alpha\beta\alpha'\beta'}^{VV+AA}
\label{D11def} \end{equation}
where ${\cal P}_{\alpha\beta\alpha'\beta'}^{VV+AA}$ is a suitably normalized
spin-projector reflecting the $VV+AA$ Dirac structure of the vertex. Details
on this projector and the notation can be found in~\cite{Donini:1999sf}.

As usual in lattice RI-MOM computations, the relevant Green 
functions~(\ref{qprop})--(\ref{5pointCF}) are evaluated
in the Landau gauge for a sequence of
sea $\{ \mu_{sea} \}$ and valence $\{ \mu_{val} \}$ 
quark mass parameters at each of the three lattice spacings 
we consider here. The bare parameters and the statistics of this computation are  
detailed in the Table~2 of ref.~\cite{Constantinou:2010gr}.
Also the lattice momenta $p_\nu$, with $p_{1,2,3} = (2\pi/L)n_{1,2,3}$, 
$p_4 = (2\pi/L)(n_4 + 1/2)$, we consider here are specified in 
eqs.~(20)--(21) of ref.~\cite{Constantinou:2010gr}. To minimise the
contributions of Lorentz non-invariant discretization effects we have
considered in our analysis only the momenta $p$ satisfying the constraint
\begin{equation}
\sum_\rho \tilde{p}_\rho^4 \, < \, 0.28 \, ( \sum_\nu \tilde{p}_\nu^2 )^2 \, ,
\qquad a\tilde{p}_\nu \equiv \sin(ap_\nu) \, .  
\label{s4cut} \end{equation} 
In the following we shall often use the quantity $\tilde{p}^2 = \sum_\nu \tilde{p}_\nu^2$. 

\subsection*{Sketch of procedure for extracting $Z_{VA+AV}^{\rm RGI}$}

Here we summarise the   
rather standard procedure we follow to extract
$Z_{VA+AV}^{\rm RGI}$, deferring the discussion of further details to  
subsequent subsections. First of all, for each $\beta$ and choice of the scale $\tilde{p}^2$, the 
condition~(\ref{VVAArencon}) is enforced at all the values of $\mu_{sea}$ and $\mu_{val}$ 
given in Table~2 of ref.~\cite{Constantinou:2010gr}. By doing so one obtains lattice approximants 
at non-zero quark mass(es) of the desired RC, $Z_{VA+AV}^{\rm RI'}(\tilde p^2; a^2\tilde p^2;\mu_{val},\mu_{sea})$ 
which are then extrapolated to $\mu_{val}=\mu_{sea}=0$~\footnote{To  
facilitate the discussion, in the notation 
for these quantities we shall distinguish, as it is customary (see e.g.~\cite{Constantinou:2010gr}),
the scale-dependence described by the RG eq's of continuum QCD (first argument)
from the one due to O($a^2\tilde{p}^2$) cutoff effects (second argument).}. 

\begin{table}
\hspace*{-0.0cm}
\begin{tabular}{cccccc}
\hline
\hline
$\beta\quad$ & $Z_{VA+AV}^{\rm RGI}$(M1)$\quad$ & $Z_{VA+AV}^{\rm RGI}$(M2)$\quad$ & 
$Z_A$(M1)$\quad$ & $Z_A$(M2)$\quad$ & $Z_V \quad$
\\
\hline
       &                           &                           &           &           &
\\
3.8    &  0.605(18)                &  0.620(13)                & 0.746(11) & 0.727(07) & 0.5816(02)
\\
       &                           &                           &           &           &
\\
3.9    &  0.623(11)                &  0.630(08)                & 0.746(06) & 0.730(03) & 0.6103(03)
\\
       &                           &                           &           &           &
\\
4.05   &  0.691(11)                &  0.698(08)                & 0.772(06) & 0.758(04) & 0.6451(03)
\\
       &                           &                           &           &           &
\\
\hline
\hline
\end{tabular}
\caption{Results for $Z_{VA+AV}^{\rm RGI}$(M1,M2) from the analysis discussed in
this appendix, together with the (best) final estimates of $Z_A$(M1,M2) and $Z_V$ from
ref.~\cite{Constantinou:2010gr}, for the three $\beta$-values of interest here.
}
\label{tab:A-RCval}
\end{table}

Improved and controlled estimates of $Z_{VA+AV}^{\rm RI'}(\tilde p^2; a^2\tilde p^2;0,0)$ 
are obtained by removing the perturbatively leading cutoff effects. Using NLO continuum QCD 
evolution~\cite{NLOevol4F,Buras:2000if}, the first argument is brought to a reference scale value, 
$\mu_0^2$.  The reliability of this step rests on the usual assumption that the scales $\tilde{p}^2$ 
and $\mu_0^2$ are large enough to make NLO-PT accurate (within statistical errors). The remaining 
$a^2\tilde{p}^2$ dependence in $Z_{VA+AV}^{\rm RI'}(\mu_0^2;a^2\tilde{p}^2;0,0)$ is to be 
regarded as a mere lattice artifact which will be taken care of by employing either the M1 or M2 
method (introduced in ref.~\cite{Constantinou:2010gr}), as briefly described below. 

In order to reduce the statistical error, the lattice approximants of the $Q_{VV+AA}$ RC 
have been averaged over equivalent pattern of Wilson parameters $(r_1,r_2,r_3,r_4)$ of valence quarks 
(namely $(1,1,1,-1)$ and $(-1,-1,-1,1)$) as well as over different lattice 
momenta corresponding to the same value of $\tilde{p}^2$. We have checked that performing
these averages before or after taking the chiral limit leads to perfectly consistent
(actually almost identical) results. The outcome of the whole analysis is conveniently
expressed in terms of $Z_{VA+AV}^{\rm RGI}$(M1) and $Z_{VA+AV}^{\rm RGI}$(M2). 
These are the RGI quantities whose values are quoted in Table~\ref{tab:A-RCval} together
with the values of the other RC's relevant for the $B_K^{\rm RGI}$ computation.

It is interesting to compare our results with those obtained 
from a 1-loop perturbative lattice computation. For instance in the case $\beta=3.9$
one gets $Z_{VA+AV}^{\rm 1-loop, RGI} = 0.422$, if the perturbative gauge coupling is set to $g_0^2/\langle P \rangle$ ($\langle P \rangle$ is the plaquette expectation value),
or $Z_{VA+AV}^{\rm 1-loop, RGI} = 0.675$, if the perturbative gauge coupling is set to the ``boosted''
value corresponding to the prescription of ref.~\cite{Aoki:2002uc}.
These numbers should be compared with the values in Table~\ref{tab:A-RCval}. 

\subsection*{Valence chiral limit}

For fixed values of $\beta$, $a^2\tilde{p}^2$ and $a\mu_{sea}$ we fit
$D_{11}$ (see eq.~(\ref{D11def})) to the ansatz
\begin{equation}
D_{11}(p;\mu_{val},\mu_{sea}) =
A(\tilde{p}^2;\mu_{\rm sea}) + B(\tilde{p}^2;\mu_{\rm sea})\mu_{val}
+ C(\tilde{p}^2;\mu_{\rm sea})(\mu_{val})^{-1} \, ,
\label{D11fit}
\end{equation}
the form of which we would like now to justify.

\begin{figure}[!ht]
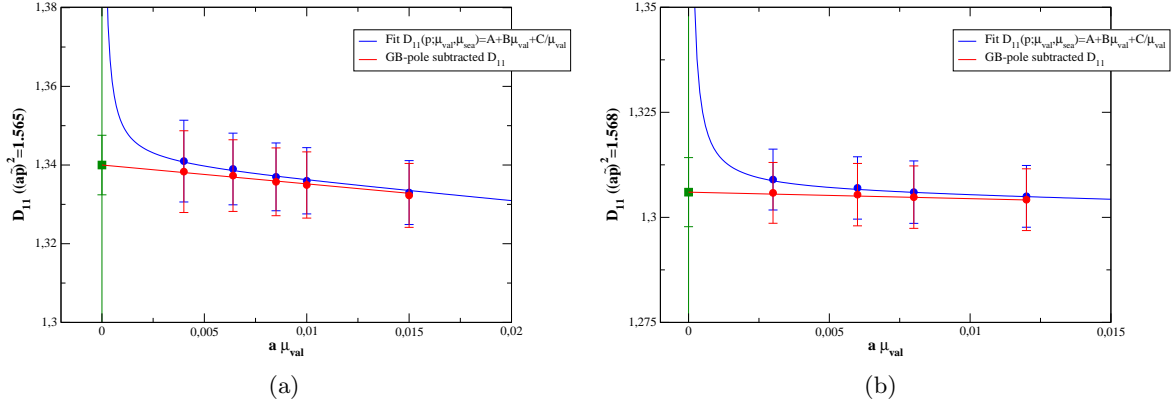

\subfigure[]{\includegraphics[scale=0.2,angle=-0]{figures_ps/fig8a.ps}}
\hspace{.2cm}
\subfigure[]{\includegraphics[scale=0.2,angle=-0]{figures_ps/fig8b.ps}}
\vskip -0.5cm
\begin{center}
\caption{GB-pole subtraction and valence chiral limit of $D_{11}$ (see eq.~(\ref{D11fit})),
         here plotted vs.\ $a\mu_{val}$, 
         for $\beta=3.9$, $a\mu_{sea}=0.0040$, $a^2\tilde{p}^2 \simeq 1.565$ (left panel)
         and $\beta=4.05$, $a\mu_{sea}=0.0030$, $a^2\tilde{p}^2 \simeq 1.568$ (right panel).
}
\label{fig:A-Valchlim}
\end{center}
\end{figure}

In general the RI-MOM approach relies on the fact that, at very large values of $p^2$ 
($\gg\Lambda_{\rm QCD}^2$), the relevant Green functions (and the associated vertices) are polynomial in
all the quark mass parameters, because non-perturbative contributions which can violate 
this property vanish in this limit. However, at finite values of $p^2$ special 
care must be taken in studying the quark mass behaviour of Green functions that 
admit one Goldstone boson (GB) intermediate states in their spectral decomposition. 
Indeed, such non-perturbative contributions to Green functions, though suppressed by a
factor $1/p^2$ for each GB pole (see below), are divergent in the chiral limit and 
must hence be disentangled
and removed.  
As explained in Appendix~B of ref.~\cite{Reyes-et-al}, this is precisely the case for the
correlator $G_{VV+AA}$ (see eq.~(\ref{5pointCF})) and the associated vertex  $D_{11}$. 

By using the LSZ reduction formulae one finds that the GB pole 
contributions relevant for (the spectral decomposition of) $D_{11}(p;\mu_{val},\mu_{sea})$ 
are of the form~\footnote{Here $q_1(p)$ denotes the four-dimensional Fourier transform 
of $q_1(x)$ with momentum $p$. As in sect.~\ref{sec:tmQCD-gen} $M^{12}$
($M^{34}$) is the mass of the lattice GB state $|P^{12}\rangle$ ($|P^{34}\rangle$).}
\begin{enumerate}
\item $\quad \dfrac{\Lambda_{\rm QCD}^2}{(M^{34})^2 p^2} 
\langle P^{34} | Q_{VV+AA}\, q_1(p) \bar{q}_2(-p) | \Omega \rangle p^2 $\, ,  
\item $\quad \dfrac{\Lambda_{\rm QCD}^2}{(M^{12})^2 p^2} 
\langle \Omega | q_3(p) \bar{q}_4(-p)\, Q_{VV+AA} | P^{12} \rangle p^2 $\, , 
\item $\quad \dfrac{\Lambda_{\rm QCD}^2}{(M^{34})^2 p^2} 
      \langle P^{34} | Q_{VV+AA} | P^{12} \rangle
       \dfrac{\Lambda_{\rm QCD}^2}{(M^{12})^2 p^2} $\, ,
\end{enumerate}
where, for simplicity, all the coefficients that are regular in the GB squared mass 
and weakly (at most logarithmically) depending on $p^2$ have been set to unity.
Furthermore immaterial (for this discussion) lattice artifact corrections are neglected. 
As discussed in sect.~\ref{sec:tmQCD-gen}, owing to the choice~(\ref{RPAR}) 
of the valence quark Wilson parameters, 
the axial current $\bar q_4 \gamma_\mu\gamma_5 q_3$ is exactly conserved 
on the lattice (only broken by soft mass terms), 
while the conservation of $\bar q_2 \gamma_\mu\gamma_5 q_1$ 
is spoiled by lattice artifacts. From the form of $Q_{VV+AA}$ 
it follows in particular that the matrix elements 
$$ \langle P^{34} | Q_{VV+AA} q_1(p) \bar{q}_2(-p) | \Omega  \rangle \, , \quad
   \langle P^{34} | Q_{VV+AA} | P^{12} \rangle $$
vanish as $(M^{34})^2 \sim (\mu_3 + \mu_4)$ in the limit $\mu_{3,4} \to 0$. 
This property can be checked by using for these matrix elements  
the soft pion theorem associated to the conservation of the lattice current
$\bar q_4 \gamma_\mu\gamma_5 q_3$ in order to reduce the GB-particle $P^{34}$. Hence
in the spectral decomposition of $D_{11}$ no terms with two GB-poles
occur. Moreover, the terms (ii) and (iii) above, which both contain one GB-pole, are strongly
suppressed by factors of the kind $a^2/p^2$ and $(1/p^2)^2$, respectively. 

In line with these arguments a very smooth dependence of $D_{11}$ on $\mu_{val}$
(or equivalently on $M_{12}^2$) is observed for all the momenta of interest in our
analysis, see Fig.~\ref{fig:A-Valchlim} for two typical examples. By fitting the $D_{11}$ 
data to the ansatz~(\ref{D11fit}) the valence chiral limit is thus safely determined. 
A practically identical result is obtained by employing a fit ansatz analogous to~(\ref{D11fit}),  
with $\mu_{val}$ substituted by $M_{12}^2$. Combining the 
valence chiral limit lattice estimator of $D_{11}$ with the corresponding approximant of $Z_q$ (the valence 
chiral extrapolation of which poses no problems~\cite{Constantinou:2010gr}) leads to reliable estimates 
of the intermediate quantities $Z_{VA+AV}^{\rm lat}(\tilde{p}^2;a^2\tilde{p}^2;0,a\mu_{sea})$.
  
\vspace{0.7cm}
\begin{figure}[!ht]
\begin{center}
\includegraphics[scale=0.25,angle=-0]{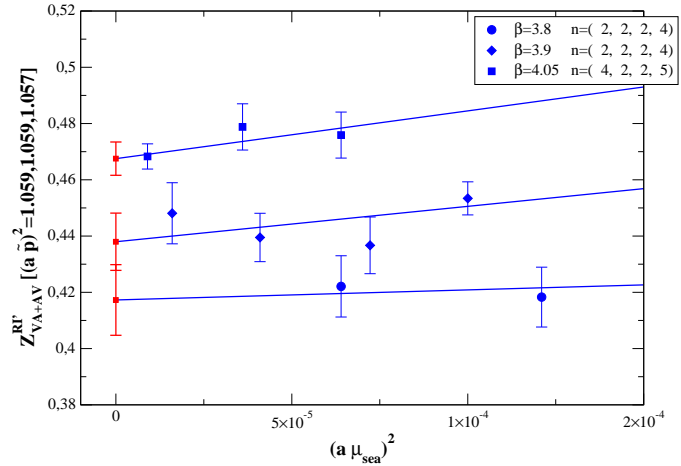}
\end{center}
\vspace{-0.2cm}
\begin{center}
\caption{The quantity $Z_{VA+AV}^{\rm RI'}(\tilde{p}^2;a^2\tilde{p}^2;0,a\mu_{sea})$, taken
 at the valence chiral limit, as a function of $a^2\mu_{sea}^2$, for a typical lattice momentum choice 
(see inset) giving $a^2\tilde{p}^2 \sim 1.06$, for the three $\beta$ values considered in this paper.
}
\label{fig:A-musea2dep}
\end{center}
\end{figure}

\subsection*{Sea chiral limit}

This limit is taken at fixed $\beta$ and $a^2\tilde{p}^2$ by fitting 
the $Z_{VA+AV}^{\rm RI'}(\mu_0^2;a^2\tilde{p}^2;0,a\mu_{sea})$-data to a polynomial of first 
order in $a^2\mu_{sea}^2$. Within our statistical errors (typically at the level of
1 to 2\%) on the data at fixed $a\mu_{sea}$, the dependence 
on $a\mu_{sea}$ is hardly significant, as one can see e.g.\ from Fig.~\ref{fig:A-musea2dep}. 
We checked that repeating the whole analysis with
$Z_{VA+AV}^{\rm RI'}(\mu_0^2;a^2\tilde{p}^2;0,a\mu_{sea})$ fitted instead to a
constant in $a\mu_{sea}$ leads to compatible results for the desired RC, but with
smaller errors and often (but not always, in particular at $\beta=3.9$) acceptable 
$\chi^2$-values. Based on these findings we conservatively decided to perform the
sea chiral extrapolation via a linear fit in $a^2\mu_{sea}^2$. 
In this way we get the RC-estimators $Z_{VA+AV}^{\rm RI'}(\mu_0^2;a^2\tilde{p}^2;0,0)$. 


\begin{figure}[!ht]
\begin{center}
\vspace{.7cm}
\includegraphics[scale=0.25,angle=-0]{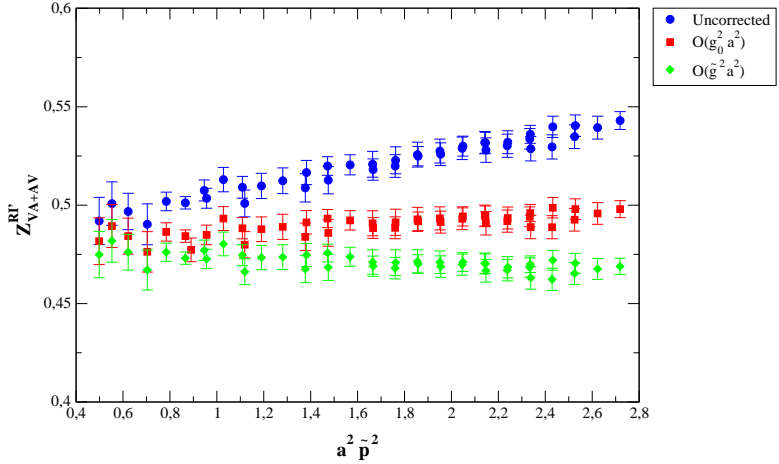}
\end{center}
\vspace{-0.2cm}
\begin{center}
\caption{The effect of subtracting from $Z_{VA+AV}^{\rm RI'}(\tilde{p}^2;a^2\tilde{p}^2;0,0)$
at $\beta=4.05$ (blue dots) the O($a^2g^2$) correction~(\ref{PTcorrZ}), setting
either $g^2=g_0^2$ (red squares) or $g^2=\tilde{g}^2$ (green diamonds).
}
\label{fig:A-Oa2PTrem}
\end{center}
\end{figure}

\subsection*{Removal of O($a^2g^2$) cutoff effects}

Improved chiral limit RC-estimators, $Z_{VA+AV}^{\rm RI'-impr}(\tilde{p}^2;a^2\tilde{p}^2)$,
are obtained  by removing from $Z_{VA+AV}^{\rm RI'}(\tilde{p}^2;a^2\tilde{p}^2;0,0)$ discretization
errors up to O($a^2g^2$), viz. 
\begin{eqnarray}
& Z_{VA+AV}^{\rm RI'-impr}(\tilde{p}^2;a^2\tilde{p}^2) = 
Z_{VA+AV}^{\rm RI'}(\tilde{p}^2;a^2\tilde{p}^2;0,0)
\nonumber \\
& + \dfrac{g^2}{16\pi^2} 
     a^2 \left[ 
     \tilde{p}^2 ( c_{D_{11}}^{(1)} + c_{D_{11}}^{(2)} \log(a^2\tilde{p}^2) ) 
     + c_{D_{11}}^{(3)} \dfrac{\sum_\rho \tilde{p}_\rho^4}{\tilde{p}^2} 
                         \right] 
\nonumber \\
& - \dfrac{g^2}{12\pi^2} 
     \, 2 \, a^2 \left[ 
     \tilde{p}^2 ( c_{q}^{(1)} + c_{q}^{(2)} \log(a^2\tilde{p}^2) ) 
     + c_{q}^{(3)} \dfrac{\sum_\rho \tilde{p}_\rho^4}{\tilde{p}^2} 
                         \right] \, .
\label{PTcorrZ} \end{eqnarray} 
The coefficients $c_{q}^{(j)}$, $j=1,2,3$ can be found in eq.~(34) of ref.~\cite{Constantinou:2010gr},
while the coefficients $c_{D_{11}}^{(j)}$, $j=1,2,3$, namely
\begin{equation}
 c_{D_{11}}^{(1)} = 2.64223 \, , \qquad c_{D_{11}}^{(2)} = -19/18 \, ,
  \qquad c_{D_{11}}^{(3)} = -2.79899 \, ,
\label{c-D11} \end{equation}
have been recently computed by evaluating the vertex $D_{11}(p)$ (eq.~(\ref{D11def})) in lattice perturbation 
theory, to 1-loop and including cutoff effects up to the second order in $a$~\cite{HP4ferm}. 
Here we just recall that in lattice perturbation theory the vertex $D_{11}(p)$ can be expressed in the form
\begin{eqnarray}
& D_{11}(p)^{\rm pert} = 1 +  \dfrac{g^2}{16\pi^2} \left[ b_{D_{11}}^{(1)} + b_{D_{11}}^{(2)} \log(a^2\tilde{p}^2) ) \right] 
\nonumber \\
& + \dfrac{g^2}{16\pi^2} a^2 \left[ 
     \tilde{p}^2 ( c_{D_{11}}^{(1)} + c_{D_{11}}^{(2)} \log(a^2\tilde{p}^2) ) 
     + c_{D_{11}}^{(3)} \dfrac{\sum_\rho \tilde{p}_\rho^4}{\tilde{p}^2} 
                         \right]  + {\rm O}(a^4g^2, g^4)
\label{expr-D11} \end{eqnarray}
and refer to~\cite{HP4ferm} for the values of the (here irrelevant) coefficients $b_{D_{11}}^{(1,2)}$ and all details. 
The form of eq.~(\ref{PTcorrZ}) follows directly from eqs.~(\ref{VVAArencon}) and ~(\ref{expr-D11}) above, 
as well as from eq.~(32) of~\cite{Constantinou:2010gr}, where the perturbative expression of 
$\Sigma_1(p)=Z_q(p)$, including the O($a^2g^2$) corrections, is given.

In the numerical evaluation of the perturbative correction in eq.~(\ref{PTcorrZ}) $g^2$ 
was taken as the simple boosted coupling $\tilde{g}^2 \equiv g_0^2/\langle P \rangle$,
where the average plaquette $\langle P \rangle$ is computed 
non-perturbatively. The values of $\langle P \rangle$ we employed here are
$[0.5689,0.5825,0.6014]$ for $\beta=[3.8,3,9,4.05]$, respectively. The (nice) effect 
of the~(\ref{PTcorrZ}) correction in removing the unwanted $a^2\tilde p^2$ dependence 
is illustrated, in the case $\beta=4.05$, in Fig.~\ref{fig:A-Oa2PTrem}. In the figure the uncorrected 
values of $Z_{VA+AV}^{\rm RI'}(\tilde{p}^2;a^2\tilde{p}^2;0,0)$ are compared with
the values of $Z_{VA+AV}^{\rm RI'-impr}(\tilde{p}^2;a^2\tilde{p}^2)$ obtained by
setting either $g^2=g_0^2=6/\beta$ or (as we chose in the end) $g^2=\tilde{g}^2$.

\subsection*{Absence of wrong chirality mixings}

\begin{figure}[!ht]
\begin{center}
{\includegraphics[scale=0.39,angle=-0]{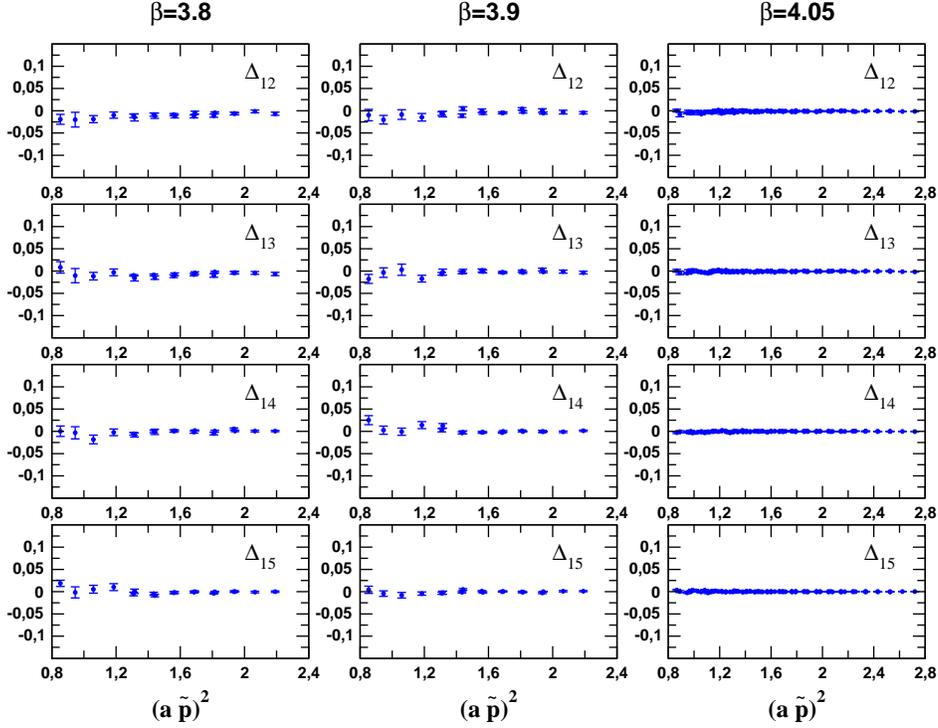}}
\end{center}
\vspace{-0.5cm}
\begin{center}
\caption{The behaviour of the mixing coefficients $\Delta_{1j}$, $j=2,3,4,5$ as function of $a^2\tilde p^2$ 
for $\beta =3.8$, $\beta =3.9$ and $\beta =4.05$, respectively.
}
\label{fig:DELTAS}
\end{center}
\end{figure}

According to the analysis of ref.~\cite{Frezzotti:2004wz}, at maximal twist wrong 
chirality mixings can affect the renormalization of $Q_{VV+AA}$ only at order $a^2$ 
(or higher). The relation between the renormalized lattice and bare operators has then the form 
\begin{equation}
\hat Q_{VV+AA}^{\rm RI'}=Z_{VA+AV}^{\rm RI'}\big{[}Q_{VV+AA}+\sum_{j=2}^5 \Delta_{1J}Q_j+\dots \big{]}\, ,
\label{WCM}
\end{equation}
where (using the operator basis of ref.~\cite{Donini:1999sf}) we have 
\begin{eqnarray}
&&Q_2=  2 \left\{ [\bar q_1 \gamma_\mu q_2][\bar q_3 \gamma_\mu q_4] -
[\bar q_1 \gamma_\mu \gamma_5 q_2][\bar q_3 \gamma_\mu \gamma_5 q_4] + (q_2 \leftrightarrow q_4) \right\}\, ,\nonumber \\
&&Q_3= 2 \left\{ [\bar q_1 q_2][\bar q_3 q_4] -
[\bar q_1 \gamma_5 q_2][\bar q_3 \gamma_5 q_4] + (q_2 \leftrightarrow q_4) \right\} \, ,\nonumber \\
&&Q_4=  2 \left\{ [\bar q_1 q_2][\bar q_3 q_4] +
[\bar q_1 \gamma_5 q_2][\bar q_3 \gamma_5 q_4] + (q_2 \leftrightarrow q_4) \right\}\, ,\nonumber \\
&&Q_5=  2 \left\{ [\bar q_1  \sigma_{\mu\nu} q_2][\bar q_3 \sigma_{\mu\nu} q_4] +
(q_2 \leftrightarrow q_4) \right\}\, .\label{QJ}
\end{eqnarray}
In eq.~(\ref{WCM}) the mixing coefficients $\Delta_{1j}$ are O($a^2$) quantities  and $\dots$ stand for higher 
order lattice corrections. In Fig.~\ref{fig:DELTAS} we plot $\Delta_{1j}$, $j=2,3,4,5$ as function of $a^2\tilde p^2$ 
for $\beta =3.8$, $\beta =3.9$ and $\beta =4.05$. It is clearly seen that in all cases the mixing coefficients are zero 
within errors and that this is systematically more so as $\beta$ increases.   

\vspace{0.9cm}
 
\subsection*{Final estimates from M1 and M2 methods}

The first step is to convert $Z_{VA+AV}^{\rm RI'-impr}(\tilde{p}^2;a^2\tilde{p}^2)$ to 
$Z_{VA+AV}^{\rm RI'-impr}(\mu_0^2;a^2\tilde{p}^2)$ by using the known formula for the NLO 
running~\cite{NLOevol4F,Buras:2000if}. This step is necessary in order to disentangle the O($a^2\tilde{p}^2$) 
cutoff effects from the genuine continuum $p^2$ dependence, but the actual value of $\mu_0$ has no 
impact on the RGI result for the RC. As customary, we take $\mu_0=1/a(\beta)$ for each value of $\beta$, 
namely $1/a(3.8)=2.0$~GeV, $1/a(3.9)=2.3$~GeV and $1/a(4.05)=2.9$~GeV. 

\begin{figure}[!ht]
\vspace{0.4cm}
\begin{center}
\includegraphics[scale=0.20,angle=-0]{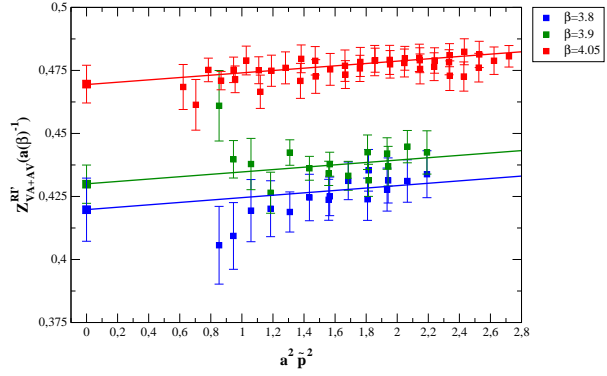}
\end{center}
\begin{center}
\caption{$Z_{VA+AV}^{\rm RI'-impr}(a(\beta)^{-2};a^2\tilde{p}^2)$ as a function of 
         $a^2\tilde{p}^2$ for the three $\beta$ values considered in this paper.
         The three straight lines represent the simultaneous linear fit
         to the $Z_{VA+AV}^{\rm RI'-impr}(a(\beta)^{-2};a^2\tilde{p}^2)$ 
         data in the interval $1.0 \leq a^2\tilde{p}^2 \leq 2.2$ at the three $\beta$'s. 
        }
\label{fig:A-p2dep}
\end{center}
\end{figure}
The method M1 consists in fitting $Z_{VA+AV}^{\rm RI'-impr}(\mu_0^{2};a^2\tilde{p}^2)$ to the ansatz
\begin{equation}
 Z_{VA+AV}^{\rm RI'-impr}(\mu_0^{2};a^2\tilde{p}^2) 
  = Z_{VA+AV}^{\rm RI'}(\mu_0^{2}) + \lambda_{VA+AV} a^2\tilde{p}^2
\label{M1fit}
\end{equation} 
in the large momentum region, $1.0 \leq a^2\tilde{p}^2 \leq 2.2$. 
As expected, the slope $\lambda_{VA+AV}$ exhibits a very mild gauge coupling dependence 
which (unlike the case of quark bilinear RC's) is not significant
within our current statistical errors. We thus treat $\lambda_{VA+AV}$ as a  
$\beta$-independent quantity. According to the ansatz~(\ref{M1fit}) a simultaneous 
linear extrapolation to $a^2\tilde{p}^2=0$ at the three $\beta$ values of interest was performed.
The extrapolated values, $Z_{VA+AV}^{\rm RI'}(\mu_0^{2})$, are finally used to
evaluate via eq.~(\ref{Z4F-RGI}) to NLO accuracy the quantities $Z_{VA+AV}^{\rm RGI}$(M1) 
quoted in Table~\ref{tab:A-RCval}.

The numbers for $Z_{VA+AV}^{\rm RI'-impr}(\mu_0^{2};a^2\tilde{p}^2)$ at the three 
$\beta$'s and the lines of simultaneous best linear fit in $a^2\tilde{p}^2$ are shown in 
Fig.~\ref{fig:A-p2dep}. We recall that in this figure, as in  
all  other figures presented in this 
Appendix, only data points corresponding to the momenta $p$ satisfying the constraint~(\ref{s4cut}) 
appear. As we said above, data that refer to momenta $p$ violating the condition~(\ref{s4cut}) are
never used in our RC analysis.  

When the method M2 is used, we separately average at each $\beta$ the values of 
$Z_{VA+AV}^{\rm RI'-impr}(\mu_0^2;a^2\tilde{p}^2)$ over a narrow interval of 
momenta  
where data are 
 linear in $a^2\tilde{p}^2$. In physical units we take the same momentum interval 
for all $\beta$'s,  i.e.  $\tilde{p}^2 \in [8.0,9.5]$~GeV$^2$.


\vspace{1.2cm}

\renewcommand{\thesection}{B}
\section*{Appendix B -- Chiral and continuum extrapolations}
\label{sec:APP-B}

Here we give some details and discuss a few technical aspects of the auxiliary 
analyses that were performed, as discussed in sect.~\ref{sec:FREB}, to estimate
the systematic error on our $B_K^{\rm RGI}$ computation. 

\begin{figure}[!ht]
\begin{center}
\includegraphics[scale=0.70,angle=-0]{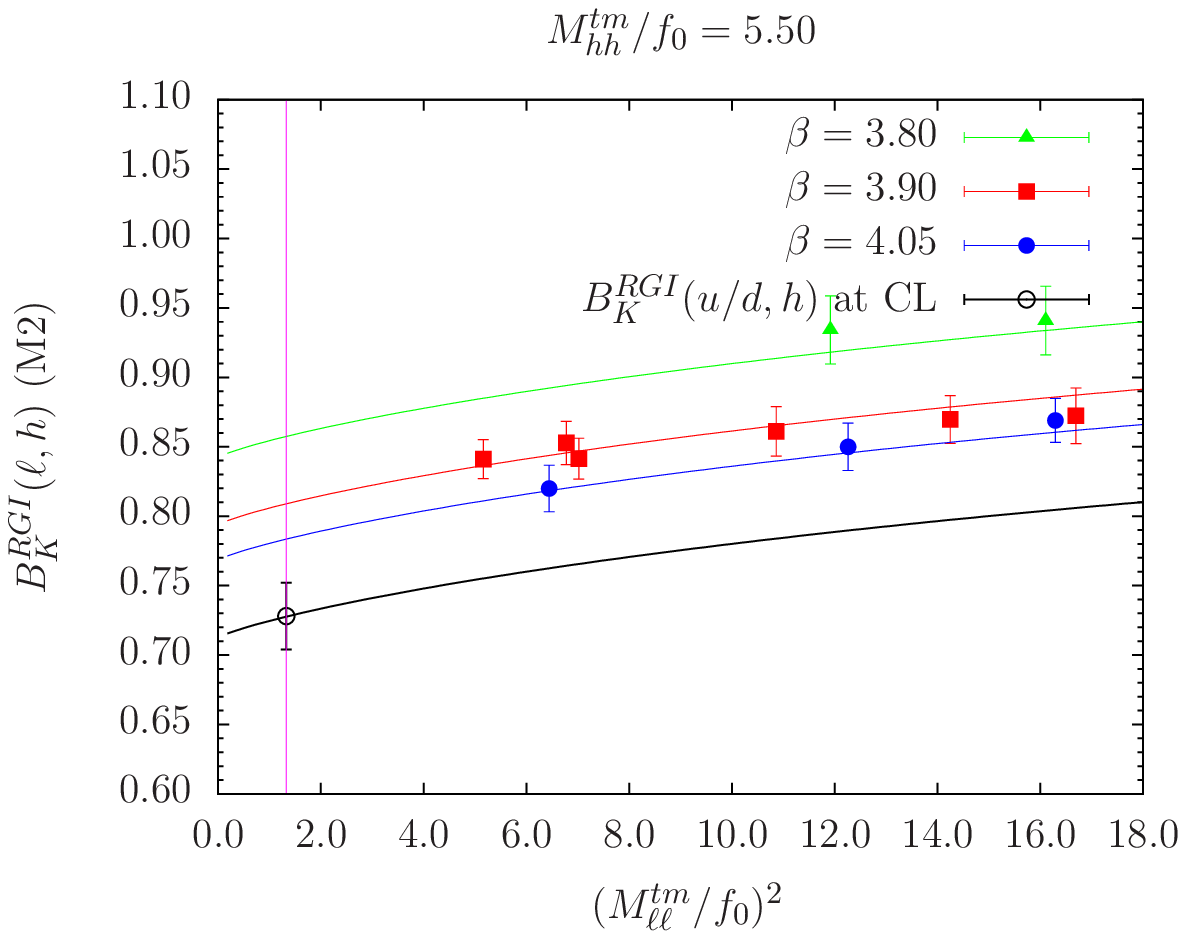}
\end{center}
\vspace{-0.2cm}
\begin{center}
\caption{Combined chiral and continuum fits according to the ansatz~(\ref{BCHIR2-mesmas})
yielding $B_K^{\rm RGI}$ at $\hat\mu_h$  
corresponding to  $M_{hh}^{\rm tm}/f_0 = 5.50$. 
The label M2 
refers to the RI-MOM procedure adopted for the computation
of $Z_{VA+AV}^{\rm RGI}$ and $Z_A$. 
}
\label{fig:BKM2M2_cochfit_mesmas}
\end{center}
\end{figure}

\begin{figure}[!ht]
\begin{center}
\includegraphics[scale=0.70,angle=-0]{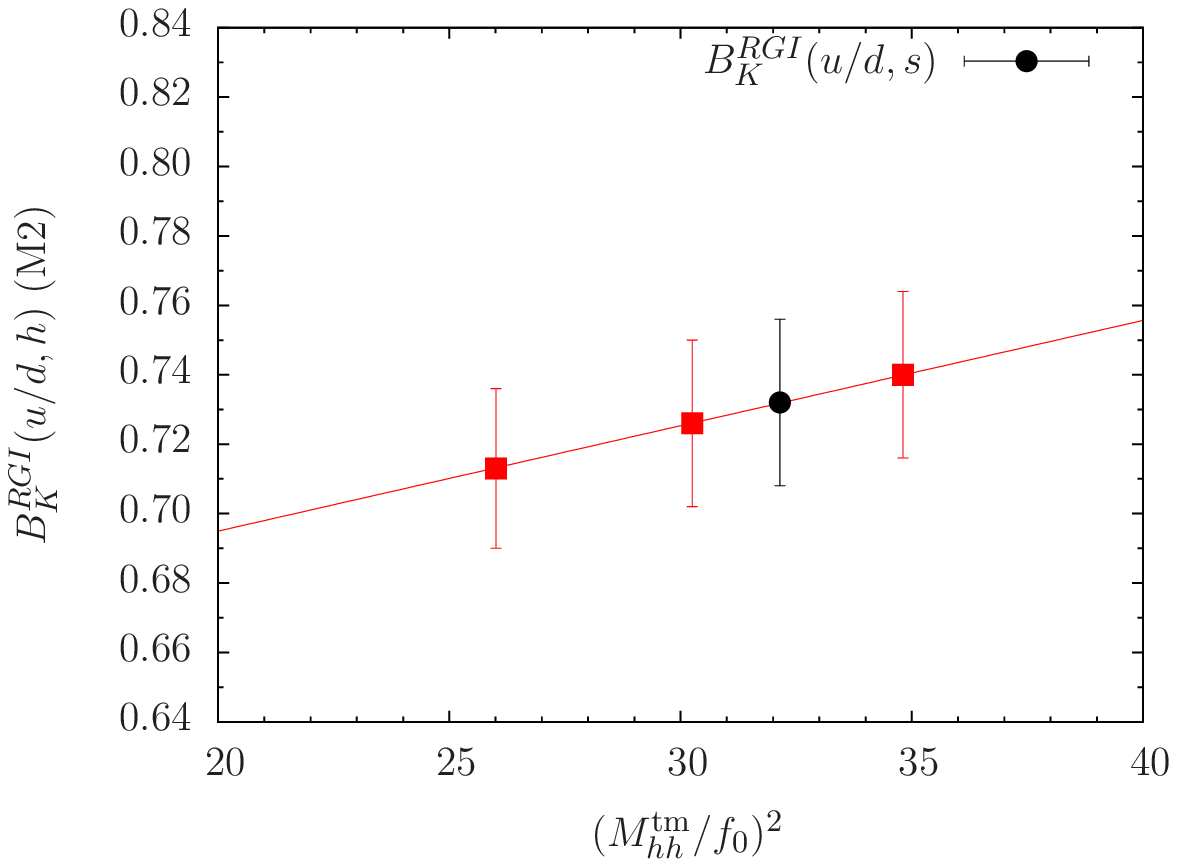}
\end{center}
\vskip -0.2cm
\begin{center}
\caption{$B_K^{\rm RGI}$ at the ``physical'' $u/d$ quark mass as a function of $(M_{hh}^{\rm tm}/f_0)^2$. 
Squares represent the values computed as explained in the text at $M_{hh} = (5.10,5.50,5.90)f_0$. 
The interpolated value at the physical point $(M_{hh}^{\rm tm})^2 = 2M_K^2 - M_\pi^2$ 
is indicated with a full dot. The label M2 has the same meaning as in Fig.~\ref{fig:BKM2M2_cochfit_mesmas}.
}
\label{fig:BKM2M2_Interp_mesmas}
\end{center}
\end{figure}

In order to evaluate the systematic errors affecting the extrapolation in the $u/d$-quark mass, 
which was performed  
together with that of the continuum limit, we also tried an
analysis where the role of the renormalized quark masses, $\hat{\mu}_\ell$
and $\hat\mu_h$, is played by  
$(M_{\ell\ell}^{\rm tm})^2$
and $(M_{hh}^{\rm tm})^2$  i.e. the masses of the pseudoscalar mesons
 made out of two mass-degenerate 
quarks (regularized with opposite Wilson parameters, as  
implied by the label ``tm''). 
The fit ansatz~(\ref{BCHIR2}) was then replaced by the following one
\begin{eqnarray}
&& B_{K,{\rm lat}}^{\rm RGI}(M_{\ell\ell}^{\rm tm},M_{hh}^{\rm tm}) = \label{BCHIR2-mesmas} \\
&& B_{K,\chi}^{\prime {\rm RGI}}(M_{hh}^{\rm tm})
\left[1 +b^\prime(M_{hh}^{\rm tm})  \dfrac{(M_{\ell\ell}^{\rm tm})^2}{f_0^2}-
\dfrac{(M_{\ell\ell}^{\rm tm})^2}{32\pi^2 f_0^2}
\log \dfrac{(M_{\ell\ell}^{\rm tm})^2}{16\pi^2 f_0^2} \right] + a^2 f_0^2 D_B^\prime(M_{hh}^{\rm tm}) \, ,
\nonumber 
\end{eqnarray}
where $B_{K,\chi}^{\prime {\rm RGI}}(M_{hh}^{\rm tm})$, $b^\prime(M_{hh}^{\rm tm})$ and
$D_B^\prime(M_{hh}^{\rm tm})$ are the three fit parameters.  
For all values of $\beta$ and $M_{\ell\ell}^{\rm tm}$,  data for 
$B_{K,{\rm lat}}^{\rm RGI}(M_{\ell\ell}^{\rm tm},M_{hh}^{\rm tm})$ have been first 
interpolated in $(M_{hh}^{\rm tm})^2$~\footnote{The interpolation was based on 
global fits linear in $(M_{hh}^{\rm tm})^2$.  
 The fit quality was
good in all cases and the statistical error on the interpolation results has been 
estimated by a standard bootstrap procedure.} in order to obtain
the RGI bag parameter at three reference values of $M_{hh}^{\rm tm}$, namely
$5.10 f_0$, $5.50 f_0$, $5.90 f_0$. 
Then three continuum and chiral fits for $B_{K,{\rm lat}}^{\rm RGI}$ at these three
reference values of $M_{hh}^{\rm tm}$ were performed, based on the 
ansatz~(\ref{BCHIR2-mesmas}). The fits turned out to be of good quality
and yielded continuum limit results for $B_K^{\rm RGI}(M_\pi,M_{hh})$ at 
$M_\pi=135$~MeV and $M_{hh} = (5.10,5.50,5.90)f_0$, respectively. For 
illustration we report in Fig.~\ref{fig:BKM2M2_cochfit_mesmas} the results for the case 
of $M_{hh}^{\rm tm} = 5.50 f_0$. The extrapolated value, $B_K^{\rm RGI}(M_\pi,5.50 f_0)$, 
is represented by an open circle in the figure. As one can see from Fig.~\ref{fig:BKM2M2_Interp_mesmas}, 
the dependence of these extrapolated results on $M_{hh}^2$ is very mild and hardly significant compared to the 
4\%-level of our statistical errors. In this situation the interpolation of $B_K^{\rm RGI}(M_\pi,M_{hh})$ 
in $(M_{hh}^{\rm tm})^2$ to the physical kaon mass (which we identified as the point where 
$(M_{hh}^{\rm tm})^2 = 2M_K^2 - M_\pi^2$) poses no particular problem (see Fig.~\ref{fig:BKM2M2_Interp_mesmas}),
and yields $B_K^{RGI}=0.732(24)$.  
 Notice that owing to the  
 weak dependence of $B_K^{\rm RGI}(M_\pi,M_{hh})$ on $(M_{hh}^{\rm tm})^2$, the
approximation induced by the use of the LO SU(3)-$\chi$PT  formula 
to relate $(M_{hh}^{\rm tm})^2$ to the physical value of the $2M_K^2 - M_\pi^2$ mass 
difference turns out to have a completely negligible impact on the final result.

\vskip 0.0cm

With the purpose of estimating the systematic uncertainty involved into the
extrapolation of our $B_K$ data to the pion (or $u/d$-quark) physical point,
we also studied the effect of performing the chiral fit assuming a first or
second order polynomial dependence on $\mu_\ell$ (or $(M_{\ell\ell}^{\rm tm})^2$). 
The analysis of the $\hat\mu_h$ and $\hat\mu_\ell$ quark mass dependence, 
described in sect.~\ref{sec:CCE}, was repeated  
with the
chiral and continuum fit ansatz~(\ref{BCHIR2})  
replaced by 
\begin{equation}
B_{K,{\rm lat}}^{\rm RGI}(\hat\mu_\ell,\hat\mu_s) = 
B_{K,\chi}^{\rm RGI}(\hat\mu_s) + C_1(\hat\mu_s) \dfrac{2\hat B_0\hat\mu_{\ell}}{f_0^2}
+ C_2(\hat\mu_s) \left( \dfrac{2\hat B_0\hat\mu_{\ell}}{f_0^2} \right)^2 +
a^2 f_0^2 C_L(\hat\mu_s)\, ,
\label{BCHIR2-poly} \end{equation}
while always taking $\hat\mu_s = 92(5)$~MeV. In the case where only a linear dependence
on $\hat\mu_\ell$ is assumed, we set $C_2(\hat\mu_s) \equiv 0$. 
Besides $B_{K,\chi}^{\rm RGI}(\hat\mu_s)$,
the other fit parameters in the formula~(\ref{BCHIR2-poly}) are
$C_L(\hat\mu_s)$,  $C_1(\hat\mu_s)$ and (only for the case of
a second order polynomial) $C_2(\hat\mu_s)$.

\begin{figure}[!h]
\begin{center}
\includegraphics[scale=0.68,angle=-0]{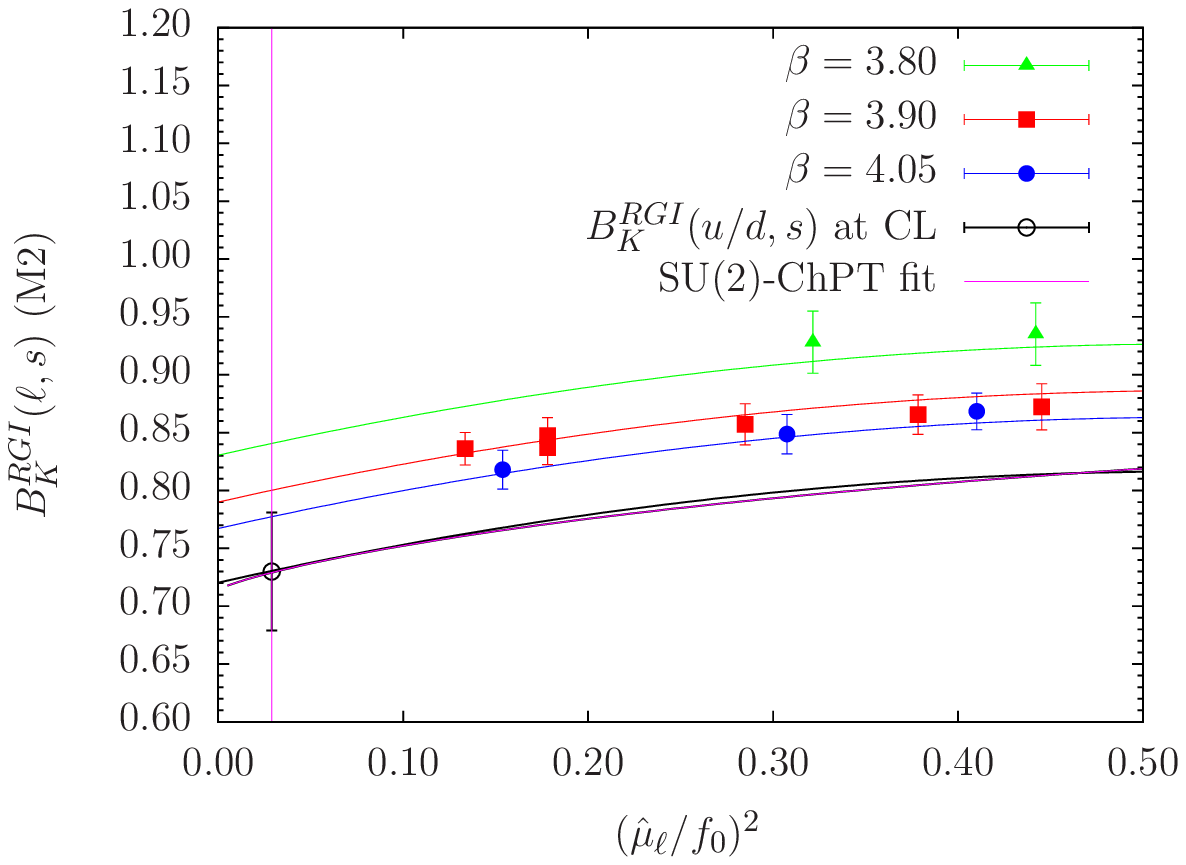}
\end{center}
\vskip -0.2cm
\begin{center}
\caption{Combined chiral and continuum fits $B_K^{\rm RGI}$
         according to the ansatz~(\ref{BCHIR2-poly}).
The label M2 has the same meaning as in Fig.~\ref{fig:BKM2M2_cochfit_mesmas}.
For comparison  
the continuum limit  curve, corresponding to the ansatz~(\ref{BCHIR2}) based on
SU(2) $\chi$PT  is also shown.
}
\label{fig:BKM2M2_cochfit_poly}
\end{center}
\end{figure}

In the case  
of an assumed  quadratic dependence of  
$B_{K,{\rm lat}}^{\rm RGI}$ data on $\hat\mu_\ell$, 
in the continuum limit and at the physical $u/d$-quark mass point 
one finds $B_K^{\rm RGI}=0.730(51)$.
The central value of this determination is perfectly in line with all 
our previous evaluations, only the associated statistical error is somewhat larger
because here we have been dealing with a four (instead of a three) parameter fit.
The good quality of the fit is well illustrated by Fig.~\ref{fig:BKM2M2_cochfit_poly}. 
A good fit yielding $B_K^{\rm RGI}=0.750(26)$ is also obtained if just
a simple linear dependence of the data on $\hat\mu_\ell$ is assumed.
Not unexpectedly (given the distribution of data points) the central value
of $B_K^{\rm RGI}$ in this case turns out to be slightly larger than in chiral fits 
that are based on SU(2) $\chi$PT or an ansatz allowing for some curvature in the 
$\hat\mu_\ell$ behaviour. This effect, which is hardly larger than one (statistical) 
standard deviation, is taken into account in our systematic error budget (see sect.~\ref{sec:FREB}). 

\vskip 0.0cm

\vskip 0.0cm
\begin{figure}[!h]
\begin{center}
\includegraphics[scale=0.68,angle=-0]{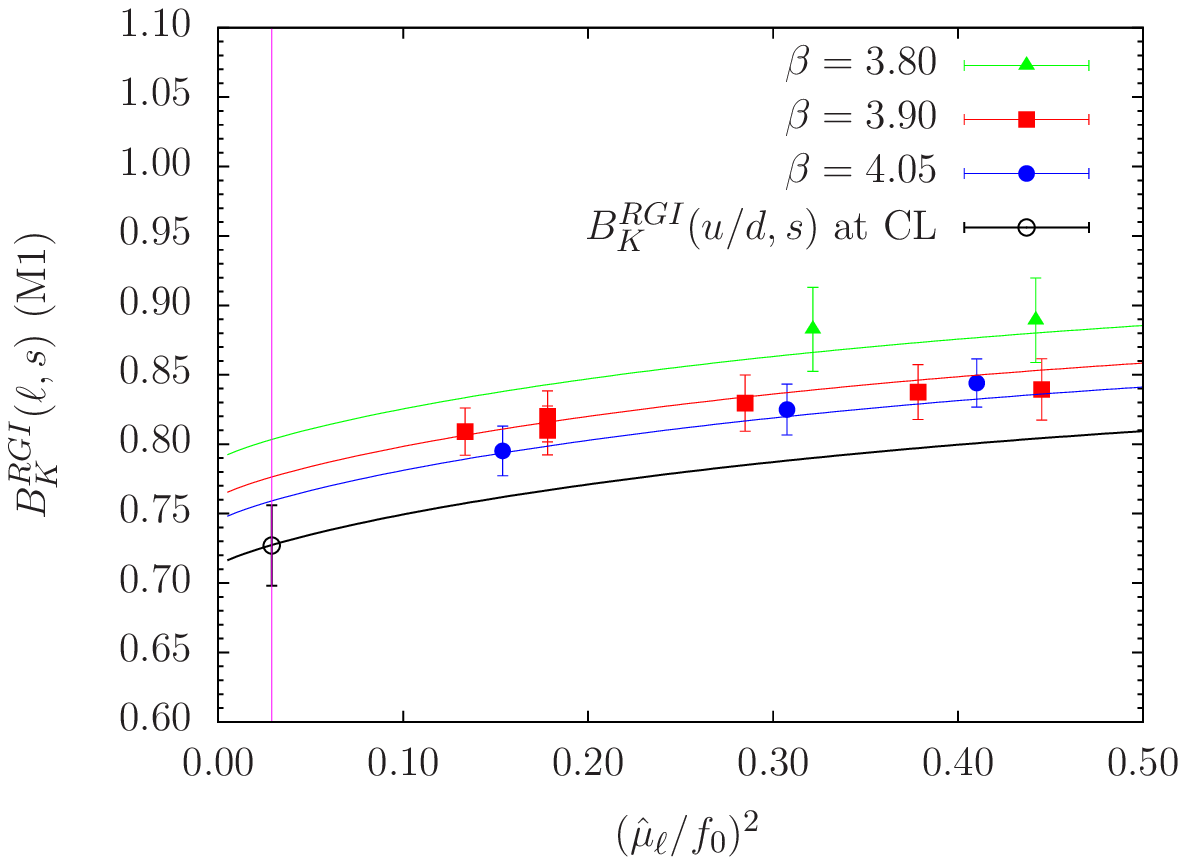}
\end{center}
\vskip -0.2cm
\begin{center}
\caption{Combined chiral and continuum fits of $B_K^{\rm RGI}$ 
         according to the ansatz~(\ref{BCHIR2}), for the case
         where the M1 method is adopted for the RI-MOM computation of
         both $Z_{VA+AV}^{\rm RGI}$ and $Z_A$ -- as implied by the 
         label M1 in the ordinate. 
}
\label{fig:BKM1M1_cochfit}
\end{center}
\end{figure}

For completeness we also show
in Fig.~\ref{fig:BKM1M1_cochfit}, the chiral and continuum
extrapolation of our $B_K^{\rm RGI}$ data
when the method M1 instead of M2 (see ref.~\cite{Constantinou:2010gr}
and Appendix~A) is employed in the RI-MOM computation of
$Z_A$ and $Z_{VA+AV}^{\rm RGI}$. By an analysis
otherwise identical to that discussed in sect.~\ref{sec:CCE}
we obtain in the continuum limit and at the physical pion mass
point $B_K^{\rm RGI} = 0.727(30)$. As
expected
from the RC values reported in Table~\ref{tab:A-RCval}, in the
present case the cutoff effects at finite lattice spacing are somewhat smaller
but the statistical errors a bit larger
compared to
the case of Fig.~\ref{fig:BKM2M2_cochfit}.

As stated in sect.~\ref{sec:FREB}, the effect of replacing with $a/r_0$ the scaling 
variable $af_0$ to build dimensionless quantities in intermediate
steps of our analyses and measure the distance from the continuum limit, 
is as small as 0.004-0.005. The curves illustrating our 
interpolation to the $s$-quark mass point and extrapolation to the
$u/d$-quark point and continuum limit in the case where $a/r_0$ is
used as scaling variable look almost identical (within 
graphical resolution) to those we presented above and are hence not shown.

Other analyses were also performed corresponding to different  
variants of our reference analysis, discussed in sect.~\ref{sec:FREB}. In all
cases the values of $B_K^{\rm RGI}$ at the physical point were found to 
be very close to the value quoted in eq.~(\ref{BK-f0M2M2}), with a spread that 
is well covered by our final estimate, $\pm 0.017$, of the systematic error.

\bibliography{}

\end{document}